\documentclass[aip,reprint,amsmath,amssymb]{revtex4-1}

\usepackage{graphicx}
\usepackage{dcolumn}
\usepackage{float}
\usepackage{bm}
\usepackage{amssymb}

\begin{document}


\title{Polymer segregation under confinement: Free energy calculations and segregation dynamics simulations}

\author{James M. Polson and Logan G. Montgomery}
\affiliation{%
Department of Physics, University of Prince Edward Island, 550 University Ave.,
Charlottetown, Prince Edward Island, C1A 4P3, Canada
}%

\date{\today}

\begin{abstract}
Monte Carlo simulations are used to study the behavior of two polymers under confinement
in a cylindrical tube. Each polymer is modeled as a chain of hard spheres.  We measure the free 
energy of the system, $F$, as a function of the distance between the centers of mass of the
polymers, $\lambda$, and examine the effects on the free energy functions of varying
the channel diameter $D$ and length $L$, as well as the polymer length $N$ and bending rigidity $\kappa$.  
For infinitely long cylinders, $F$ is a maximum at $\lambda=0$, and decreases with 
$\lambda$ until the polymers are no longer in contact.  For flexible chains ($\kappa=0$), 
the polymers overlap along the cylinder for low $\lambda$, while above some critical value 
of $\lambda$ they are longitudinally compressed and non-overlapping while still in contact. 
We find that the free energy barrier height, $\Delta F\equiv F(0)-F(\infty)$, scales as 
$\Delta F/k_{\rm B}T \sim ND^{-1.93\pm 0.01}$, 
for $N\leq 200$ and $D\leq 9\sigma$, where $\sigma$ is the monomer diameter.  In addition, 
the overlap free energy appears to scale as $F/k_{\rm B}T=Nf(\lambda/N;D)$ for sufficiently large $N$, 
where $f$ is a function parameterized by the cylinder diameter $D$.
For channels of finite length, the free energy barrier height
increases with increasing confinement aspect ratio $L/D$ at fixed volume fraction 
$\phi$, and it decreases with increasing $\phi$ at fixed $L/D$.  Increasing the polymer bending 
rigidity $\kappa$ monotonically reduces the overlap free energy.  For strongly confined systems, 
where the chain persistence length $P$ satisfies $D\ll P$, $F$ varies linearly with $\lambda$ with 
a slope that scales as $F'(\lambda)\sim -k_{\rm B}T D^{-\beta} P^{-\alpha}$, where $\beta\approx 2$ and 
$\alpha\approx 0.37$ for $N=200$ chains.  These exponent values deviate slightly from those 
predicted using a simple model, possibly due to insufficiently satisfying the conditions defining 
the Odijk regime.  Finally, we use Monte Carlo dynamics simulations to examine polymer segregation 
dynamics for fully flexible chains and observe segregation rates that decrease with decreasing entropic 
force magnitude, $f\equiv |dF/d\lambda|$. For both infinite-length and finite-length channels, the 
polymers are not conformationally relaxed at later times during segregation. 

\end{abstract}

\maketitle

\section{Introduction}
\label{sec:intro}

When two self-avoiding polymer chains overlap, each experiences a reduction in 
conformational entropy.  The resulting increase in free energy at maximum overlap approaches 
a finite value with increasing polymer length, and the polymer coils experience a 
very soft effective potential that allows for a high degree of mutual penetration.%
\cite{grosberg1982polymeric,dautenhahn1994monte} When the polymers are confined to 
a cylindrical space, the overlap free energy can be significantly affected. 
Under the right confinement conditions, it can be sufficiently high to cause segregation
of the polymers.\cite{daoud1977statistics,jun2006entropy} Such entropy-driven demixing of confined polymers 
may be relevant for some biological processes. Notably, it has been proposed to be the dominant 
mechanism for chromosome separation in bacteria, with proteins identified as segregation factors serving 
mainly to create the right conditions for entropy-driven segregation.%
\cite{jun2006entropy,jun2010entropy,youngren2014multifork}
While several recent studies have challenged this view,\cite{lechat2012let,%
yazdi2012variation,kuwada2013mapping,diventura2013chromosome,junier2014polymer}
an understanding of bacterial chromosome separation remains elusive and the importance of entropy
as a driving force is currently not clear.\cite{reyes2012chromosome,wang2013organization}

Computer simulation studies have provided insight into the segregation of polymers
under confinement.\cite{jun2006entropy,teraoka2004computer,jun2007confined,arnold2007time,jacobsen2010demixing,%
jung2010overlapping,jung2012ring,jung2012intrachain,liu2012segregation,dorier2013modelling,racko2013segregation,%
shin2014mixing,milchev2014arm} Most studies have considered systems comprised of flexible linear polymers,\cite{jun2006entropy,%
teraoka2004computer,jun2007confined,arnold2007time,jacobsen2010demixing,jung2010overlapping,jung2012intrachain,%
liu2012segregation,racko2013segregation,milchev2014arm} though some have have examined segregation of ring polymers,%
\cite{jung2012ring,dorier2013modelling,shin2014mixing} as well as the effects of bending 
rigidity\cite{racko2013segregation} and molecular crowding.\cite{shin2014mixing} 
In the case of linear polymers, the role of cylindrical confinement in modulating the entropic repulsion
has been investigated by examining the dependence of the polymer miscibility and intrachain
ordering statistics on the confinement dimensions and chain density.\cite{jung2010overlapping,jung2012intrachain}
It was found that chains demix even outside the de~Gennes linear ordering scaling regime.
In addition, it was observed that the degree of separation increases with increasing asymmetry of 
the confinement space (i.e. more elongated cylinders) and decreases with increasing volume fraction 
of the monomers. The variation of the chain miscibility with the confinement dimensions was
used to predict the corresponding behavior for ring polymers using a scaling of the 
cylinder diameter $D\rightarrow D/\sqrt{2}$,\cite{jung2012ring} and the analysis suggests that 
{\it E. coli} chromosomes are linearly ordered and lie in the spontaneous segregation regime.%
\cite{jung2012ring,jung2012intrachain}

The miscibility of confined polymers is determined by the form of the variation of the free energy
with the degree of polymer overlap. In principle, this can be calculated using probability distributions
obtained directly from simulations. Jung {\it et al.} measured such distributions for the distance 
between the polymer centers of mass of two flexible chains and found that they were sharply peaked at 
wide separations for sufficiently large $L/D$ but were significantly broader for more symmetric 
confinement.\cite{jung2012intrachain} The latter result is consistent with the prediction that there is
no entropic cost for segregating chains in a spherical confining cavity.\cite{jun2007confined}
The results are indicative of a free energy function with a deep minimum at large $L/D$ that
becomes more shallow as the confining space becomes more symmetric. 
Polymer miscibility under confinement has also been studied using analytic forms of the free energy function 
estimated using a de~Gennes blob scaling analysis.\cite{jun2007confined,arnold2007time,liu2012segregation} 
Assuming that the polymer chain remains in conformational quasi-equilibrium, the
free energy functions can be used to predict the dynamics of polymer segregation.
Arnold and Jun used such a simple model to study segregation in infinitely long cylinders and interpret
results from Langevin simulations.\cite{arnold2007time}
Liu and Chakraborty employed an estimate of the free energy for a mean first-passage 
time analysis of the Fokker-Planck equation to study the dynamics of segregating chains in a 2-D
system.\cite{liu2012segregation} In different limiting cases, the predictions yielded a scaling
of the segregation time, $\tau$, with polymer length that was consistent with results from Monte Carlo (MC)
dynamics calculations.  However, quantitative discrepancies were appreciable, pointing to the limitations
of the simple analytical model employed to describe the free energy function.

In this study, we use MC simulations to calculate the free energy as a function of polymer separation 
for two polymer chains confined to a cylindrical tube. In order to obtain quantitatively accurate functions 
over a wide range of separations, we employ a technique\cite{frenkel2002understanding} that we have 
previously used to calculate free energy functions for polymer translocation.\cite{polson2013simulation,%
polson2013polymer,polson2014evaluating} We study the dependence of the free energy functions on 
the dimensions of the confining cylinder, the polymer length, the volume fraction 
and the polymer stiffness. To our knowledge, this is the first such systematic study 
of the overlap free energy for cylindrically confined systems.\cite{note1} The calculated free energies 
are used to test predictions of simple analytical models used in other studies.
We find that scaling arguments can account for the observed trends, though the measured
scaling exponents differ somewhat from the predicted values. 
In addition, we carry out MC dynamics 
simulations to study the segregation dynamics and interpret the results using the free energy functions. 
In the early stages of segregation, the observed dynamics are in semi-quantitative agreement 
with the predictions.  At longer times, the polymers are conformationally out of equilibrium, and 
the free energy functions are less useful for understanding the dynamics.

\section{Model}
\label{sec:model}

\begin{figure}[!ht]
\begin{center}
\vspace*{0.2in}
\includegraphics[width=0.4\textwidth]{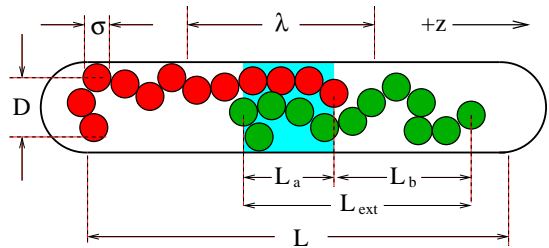}
\end{center}
\caption{Illustration of the definition of the tube diameter, $D$, and length, $L$,
for a finite length confining tube. Also shown are the following quantities measured
in the simulations: the distance between polymer centers of mass, $\lambda$; the
chain extension length, $L_{\rm ext}$; the polymer overlap distance, $L_{\rm a}$;
and the non-overlap distance, $L_{\rm b}$. These distances are measured along the 
$z$ axis.  }
\label{fig:illust}
\end{figure}

We employ a minimal model of two polymer chains confined to a cylindrical tube. Each polymer is 
modeled as a chain of $N$ hard spheres, each with a diameter of $\sigma$. The pair potential for 
non-bonded monomers is thus $u_{\rm{nb}}(r)=\infty$ for $r\leq\sigma$ and $u_{\rm{nb}}(r)=0$ for 
$r>\sigma$, where $r$ is the distance between the centers of the monomers. Pairs of bonded monomers
interact with a potential $u_{\rm{b}}(r)= 0$ if $0.9\sigma<r<1.1\sigma$ and $u_{\rm{b}}(r)= \infty$,
otherwise.  Consequently, the bond length can fluctuate slightly about its average value.
The polymers are confined to a hard cylindrical tube of diameter $D$. Thus, each monomer interacts 
with the wall of the tube with a potential $u_{\rm w}(r) = 0$ for $r<D$ and $u_{\rm w}(r) = \infty$ for $r>D$,
where $r$ is the distance of a monomer from the central axis of the cylinder. Thus, $D$ is defined to 
be the diameter of the cylindrical volume accessible to the centers of the monomers. For confining tubes 
of finite length, each end of the cylinder is capped with a hemisphere whose diameter
is equal to that of the cylinder. The length, $L$, of the capped tube is defined be
that of the cylindrical portion of the volume. This is illustrated in Fig.~\ref{fig:illust}.

Most of the simulations in this study employ fully flexible polymer chains. 
However, in some cases we consider the effects of bending stiffness on the free energy
functions. To do this, we employ a bending potential associated with each consecutive triplet
of monomers. The potential has the form, $u_{\rm bend}(\theta) = \kappa (1-\cos\theta)$.
The angle $\theta$ is defined at monomer $i$ such that $\cos\theta_i \equiv \hat{u}_{i}\cdot\hat{u}_{i+1}$,
where $\hat{u}_i$ is a normalized bond vector pointing from monomer $i-1$ to monomer $i$.
The bending constant $\kappa$ determines the stiffness of the polymer and is related
to the persistence length $P$ by $\kappa/k_{\rm B}T= P/\langle l_{\rm bond}\rangle$,
where $\langle l_{\rm bond} \rangle$ is the mean bond length.

\section{Methods}
\label{sec:methods}

Monte Carlo simulations employing the Metropolis algorithm and the self-consistent
histogram (SCH) method\cite{frenkel2002understanding} were used to calculate the free energy functions
for the polymer-nanopore model described in Section~\ref{sec:model}. The SCH method
provides an efficient means to calculate the equilibrium probability distribution
$P(\lambda)$, and thus its corresponding free energy function, $F(\lambda) = -k_{\rm B}T\ln P(\lambda)$,
where $k_{\rm B}$ is Boltzmann's constant and $T$ is temperature.
The method circumvents the problem of poor statistics that would otherwise
occur when the variation in $F$ with respect to $\lambda$ exceeds a few $k_{\rm B}T$. 
The method is similar to that used by Narros {\it et al.} to calculate 
effective potentials between overlapping polymers.\cite{narros2010influence,narros2013architecture}

To implement the SCH method, we carry out many independent simulations, each of which employs a
unique ``window potential'' of a chosen functional form.  The form of this potential is given by:
\begin{eqnarray}
{W_i(\lambda)}=\begin{cases} \infty, \hspace{8mm} \lambda<\lambda_i^{\rm min} \cr 0,
\hspace{1cm} \lambda_i^{\rm min}<\lambda<\lambda_i^{\rm max} \cr \infty, \hspace{8mm} \lambda>\lambda_i^{\rm max} \cr
\end{cases}
\label{eq:winPot}
\end{eqnarray}
where $\lambda_i^{\rm min}$ and $\lambda_i^{\rm max}$ are the limits that define the range of $\lambda$
for the $i$-th window.  Within each ``window'' of $\lambda$, a probability distribution $p_i(\lambda)$ is
calculated in the simulation. The window potential width,
$\Delta \lambda \equiv \lambda_i^{\rm max} - \lambda_i^{\rm min}$, is chosen to be sufficiently small
that the variation in $F$ does not exceed a few $k_{\rm B}T$. Adjacent windows overlap,
and the SCH algorithm uses the $p_i(\lambda)$ histograms to reconstruct the unbiased distribution,
$P(\lambda)$. The details of the histogram reconstruction algorithm are given in Ref.~\onlinecite{frenkel2002understanding},
and a description for an application to a physical system comparable to that studied here 
is presented in Ref.~\onlinecite{polson2013simulation}.

Polymer configurations were generated using translational displacements of the monomers.
The displacement in each dimension was chosen from a uniform distribution in the range 
$[-\Delta_{\rm max},\Delta_{\rm max}]$. The trial moves were accepted with a
probability $p_{\rm acc}=\min(1,e^{-\Delta E/k_{\rm B}T})$, where $\Delta E$ is
the energy difference between the trial and current states. The maximum displacement parameter 
was typically chosen to be $\Delta_{\rm max}=0.22\sigma$.  In addition to translational monomer 
displacements, we also employed standard reptation moves in simulations of systems with finite bending 
rigidity. Initial polymer configurations were generated such that $\lambda$ was within the
allowed range for a given window potential. The system was equilibrated for typically $\sim 10^7$
MC cycles, following which a production run of $\sim 10^8$ MC cycles was carried out.
During each MC cycle a move for each monomer is attempted once, on average.

The windows are chosen to overlap with half of the adjacent window, such that 
$\lambda^{\rm max}_{i} = \lambda^{\rm min}_{i+2}$.  The window widths were typically $\Delta \lambda = \sigma$
or $\Delta\lambda=2\sigma$.  The number of windows (and, therefore, the number of independent simulations) was 
determined by the range of $\lambda$. The range was given by $\lambda \in [0,\lambda_{\rm max}]$.
In the $L=\infty$ case,  the value of $\lambda_{\rm max}$ was determined by the condition that the
polymers had negligible probability of contact. This corresponds to a region where the free energy
is constant with respect to $\lambda$. As an illustration, for a system of polymers of
length $N=200$ in a cylinder of diameter $D=4\sigma$, a value of $\lambda_{\rm max}=120\sigma$ was used. This 
required simulations for 119 different window potentials for $\Delta\lambda=2\sigma$.  
In the finite-$L$ case, $\lambda_{\rm max}$ 
is limited by monomer crowding near the caps of the cylinder, and values were chosen to account for this.
For each simulation, individual probability histograms were constructed using the binning technique 
with 10 bins per histogram, $p_i(\lambda)$. 

In addition to free energy calculations, Monte Carlo dynamics simulations were used to study 
the segregation dynamics of the two confined polymers. Polymer motion was generated solely through 
random monomer displacement.  Each coordinate displacement was randomly chosen from a uniform 
distribution $[-0.22\sigma,0.22\sigma]$.  A trial move was rejected if it led to overlap with 
another nonbonded monomer, violation of the bonding constraint, or overlap with the channel wall.  
The polymer configurations were initially chosen such that $\lambda = 0$. In the case of finite-$L$ tubes,
the polymer centers of mass were  fixed near the center of the tube, as measured along the $z$ axis.  
Otherwise, initialization was carried out in the same manner as described above for the free energy 
simulations.  Following equilibration, the segregation process was examined by measuring the time 
dependence of various quantities. We simulated typically 500 -- 2000 segregation events to calculate 
averages of these time-dependent quantities.

The following quantities were measured in both the free energy simulations and the MC dynamics
simulations:
(1) $\lambda$, the distance between polymer centers of mass;
(2) $L_a$, the overlap distance of the polymers along $z$; 
(3) $L_{\rm b}$, the average length of the non-overlapping portions of the chains, measured along $z$; 
(4) $L_{\rm ext} (=L_{\rm a}+L_{\rm b})$, the average extension length of each polymer, measured along $z$.
These quantities are illustrated in Fig.~\ref{fig:illust}.
In the results presented below, lengths are measured in units of $\sigma$, the
monomer diameter. In addition, time is measured in MC cycles, where 1 MC cycle corresponds
to one attempted move per monomer, on average. For simulations of polymers with finite bending
rigidity, the bending constant $\kappa$ is measured in units of $k_{\rm B}T$. 

\section{Results}
\label{sec:results}

\subsection{Infinite length confinement cylinder}
\label{subsec:open}

\begin{figure}[!ht]
\begin{center}
\includegraphics[width=0.4\textwidth]{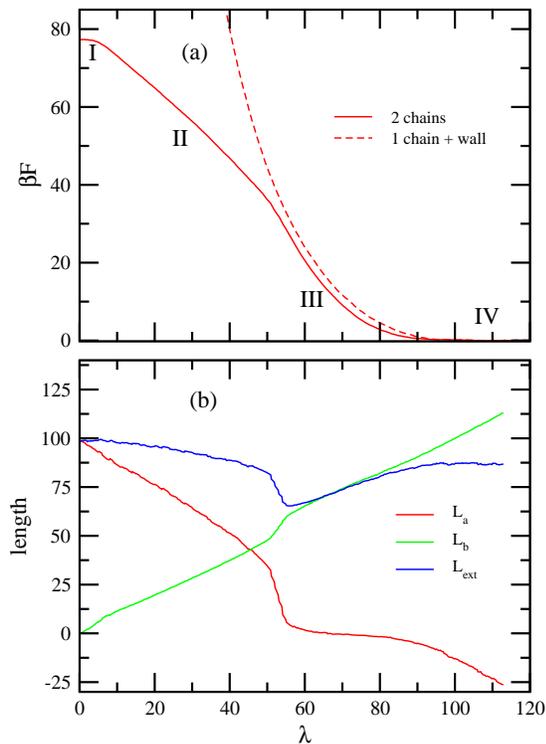}
\end{center}
\caption{(a) Free energy function for two chains of length $N=200$ in an infinitely
long cylinder of diameter $D=4$. The dashed line shows an appropriately scaled
free energy function for a single chain in a tube capped at one end with a
flat wall, as described in the text.
(b) Variation of $L_{\rm a}$, $L_{\rm b}$ and $L_{\rm ext}$ with
$\lambda$. }
\label{fig:F.Lab.N200.R2.5}
\end{figure}

We consider first the case of fully flexible polymers (i.e. $\kappa=0$) confined
to cylindrical tubes of infinite length.  Figure~\ref{fig:F.Lab.N200.R2.5}(a) shows a 
free energy function for a system of polymers of length $N=200$ confined to a tube of 
diameter $D=4$. As expected, the free energy increases as the overlap of the polymer chains 
increases, and reaches a maximum when $\lambda=0$. This is due to the reduction
in conformational entropy of overlapping chains.
There are four distinguishable regions, each of which are labeled
in the figure: (I) In the region $0 < \lambda \lesssim 5$, $F$ is approximately
constant. (II) For $5 \lesssim \lambda \lesssim 51$, $F$ decreases roughly
linearly with $\lambda$, though the curve does have a slightly negative
curvature in this region. (III) For $51 \lesssim \lambda \lesssim 90$,
$F$ continues to decrease with $\lambda$, though the curve has 
clearly positive curvature. (IV) For $\lambda \gtrsim 90$, $F$ is
constant. The general shape of the function in this figure is also observed for 
different values of $N$ and $R$, though the domain boundaries are dependent
on these parameters.

Region~I is due to an effect exhibited for nearly overlapping centers of mass
in which one polymer tends to extend farther on each side than the other, with 
the longer and shorter polymers periodically swapping identities.  Only when
$\lambda$ is sufficiently large to prevent such ``nesting'' of chains will
an increase in $\lambda$ correspond to an increase in the actual overlap
in the monomers and, thus, to a change in $F$. A similar effect was observed
recently in simulations of two overlapping polymers confined to a cylinder with 
one end of each tethered to a capped end of the tube.\cite{milchev2014arm}
The significance of the other domains is clear from the 
results shown in Fig.~\ref{fig:F.Lab.N200.R2.5}(b), which shows the average
values of $L_{\rm a}$, $L_{\rm b}$, and $L_{\rm ext}$ as a function
of $\lambda$. In region II, the polymers overlap, but the overlap distance
$L_{\rm a}$ decreases as the polymer centers of mass become
more widely separated. In a narrow region of $51 \lesssim \lambda \lesssim 55$, 
there are abrupt decreases in the polymer extension length $L_{\rm ext}$ as well as
$L_{\rm a}$, which falls to zero. Following this abrupt change,
$L_{\rm a}$ remains close to zero, while $L_{\rm ext}$ and $L_{\rm b}$ both increase with $\lambda$.
Thus, region~III corresponds to the case where the polymers do not overlap, but are in
contact and longitudinally compressed.  This compression decreases as the polymer centers of
mass separate. For $L \gtrsim 90$ in region~IV, we observe $L_{\rm a} < 0$, and the value becomes increasingly
negative with increasing polymer separation. This means that the polymers are no
longer in contact and that the distance between the nearest pair monomers on each polymer 
is increasing. In addition, $L_{\rm ext}$ is constant in this region. Thus, the separation 
distance of non-interacting polymers does not affect their conformational state, and the 
free energy does not depend on the distance.

To further understand the variation of $F$ in region~III, we have carried out the following
simulation. We measure the free energy of a single $N=200$ chain in a $D=4$ tube that is
capped at one end with a flat wall. We then measure $F$ as
a function of the distance between the center of mass of the polymer and its mirror
image on the other side of the wall. This should approximate the behavior of two chains
that are in contact, but do not overlap. For a quantitative comparison, the free energy
is scaled by a factor of 2, to account for the presence of two chains in the original system. The
scaled function is shown as a dashed curve in Fig.~\ref{fig:F.Lab.N200.R2.5}(a). The
function follows that of the two-chain system very closely in region~III, with the same
positive curvature. Thus, in region~III, the polymer chains interact as though the
interface between them forms an impenetrable wall. There is a slight shift of the solid curve 
toward lower $\lambda$, which is likely due to the flexibility and slight penetrability of
the interface in the two-chain system, in contrast to the case for a true hard wall.  
Note that the dashed curve diverges from the original two-chain function once the chains 
begin to overlap, as expected.
  
\begin{figure}[!ht]
\begin{center}
\includegraphics[width=0.4\textwidth]{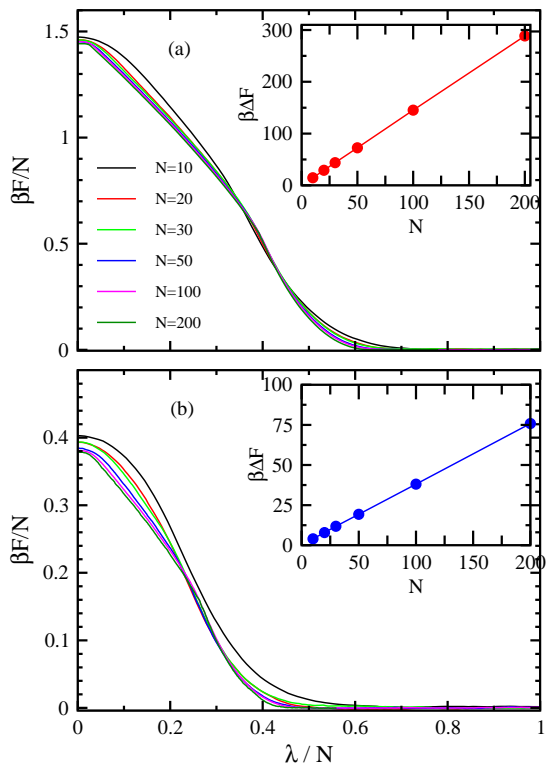}
\end{center}
\caption{(a) Scaled free energy functions for two polymer chains confined in an infinitely
long cylindrical tube of diameter $D=2$. Results are shown for several different
values of polymer length, $N$. The inset shows the variation of $\Delta F \equiv
F(0)-F(\infty)$ vs $N$. (b) As in (a), except for a tube diameter of $D=4$.
}
\label{fig:F.scale}
\end{figure}

The effects of varying the polymer chain length, $N$, on the free energy functions are
shown in Fig.~\ref{fig:F.scale}. Results are shown for $D=2$ and $D=4$ in
Figs.~~\ref{fig:F.scale}(a) and (b), respectively, and the data is scaled by $1/N$ on 
each axis. In each case, the scaled functions overlap to a considerable degree. The overlap 
is poorest for low $N$ (notably so for $N=10$). The curves appear to converge with increasing $N$,
and convergence is evidently faster for $D=2$ than for $D=4$. Thus, the results suggest that
$F/N = f(\lambda/N;D)$ for sufficiently large $N$, where $f$ is a function parameterized by
the tube diameter $D$.  The deviations for short $N$ are
likely due to chain-end effects, which are expected to be more significant for wider confinement
tubes where the chain is less stretched along $z$ and more monomers are close to the 
extrema positions of the polymer.  Confirmation of this hypothesis requires simulations for appreciably 
greater $N$, which, unfortunately, is not computationally feasible at present. The height of the free 
energy barrier, $\Delta F\equiv F(0)-F(\infty)$, varies linearly with $N$, as is evident from the scaled curves as well as
the inset of the figure. Slight deviations from linearity (not evident in the figure) occur at
low $N$ due to the finite-size effects.  Note that the rate of increase of $F$ with $N$ is greater 
for the more confined system, as expected.

A simple explanation for the apparent scaling of $F$ with $N$ is as follows. In regime~II, it
is plausible that $F\propto N_{\rm a}$, where $N_{\rm a}$ is the average number of monomers 
per polymer in the overlap region. This is consistent with the 
limiting case of maximum overlap at $\lambda=0$, where $\Delta F\propto N$ was observed.
Assuming that monomer densities are uniform (though distinct) in the overlapping and non-overlapping
regions, it is easily shown that $N_{\rm a}=N u(\lambda/N;D)$, where $u$ is a function
parameterized by $D$, which controls the polymer subchain length per monomer in the overlapping and
non-overlapping regions. Thus, $F/N = u(\lambda/N;D)$, where the proportionality constant has been
absorbed into the function $u$. In regime~III, the polymers do not overlap but are each
compressed. We employ the renormalized Flory theory of Jun {\it et al.}\cite{jun2008compression}
to estimate the variation of $F$ with $L_{\rm ext}$. Those authors note that the free energy
for a polymer of extension length $L_{\rm ext}$ is $\beta F=A L_{\rm ext}^2/(N/g)D^2
+ BD(N/g)^2/L_{\rm ext}$, where $A$ and $B$ are constants of order unity and
$g\approx (D/\sigma)^{5/3}$ is the number of monomers in a compression blob of diameter $D$.
From Fig.~\ref{fig:F.Lab.N200.R2.5}(b) we observe that $L_{\rm ext}\approx c\lambda$ in this 
regime, where $c$ is a $D$-dependent proportionality constant. It follows that 
$\beta F/N = w(\lambda/N;D)$, where $w(x;D) = (2Agc^2/D^2) x^2 + (2B/cg^2) x^{-1}$.
This expression for $F$ is expected to be valid in the regime where $L_{\rm ext}$ exceeds approximately
half of its average length.\cite{jun2008compression} Fig.~\ref{fig:F.Lab.N200.R2.5}(b) clearly
shows that this condition is satisfied in regime~III. Thus, in regimes II and III the 
observed scaling of $F$ with $N$ is predicted. The cross-over between the regimes occurs
at $\lambda^*$ determined by the condition $u(\lambda^*/N;D)=w(\lambda^*/N;D)$. Finally, 
the relative width of regime~I is expected to vanish as $N\rightarrow\infty$, and $F$ 
is of course invariant to $\lambda$ in regime~IV. Thus, we predict 
$F/N = f(\lambda/N;D)$ for sufficiently large $N$.

\begin{figure}[!ht]
\begin{center}
\includegraphics[width=0.4\textwidth]{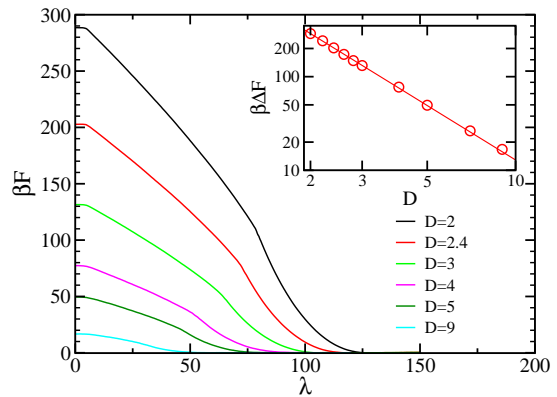}
\end{center}
\caption{Free energy functions for two polymer chains, each of length $N=200$,
confined in an infinitely long cylindrical tube.  Results are shown for several different
values of the cylinder diameter, $D$. The inset shows the variation of $\Delta F \equiv
F(0)-F(\infty)$ vs channel diameter. A scaling of $\beta\Delta F \sim D^{-1.93\pm 0.01}$ is
observed.
}
\label{fig:F.N200}
\end{figure}

Figure~\ref{fig:F.N200} shows the effect on $F(\lambda)$ of variation in the confining
tube diameter, $D$. Results are shown for polymers of length $N=200$. As expected the polymer
overlap free energy decreases with increasing diameter. This follows from the fact that an
increase in $D$ means the polymers have greater freedom in the transverse direction (i.e. $x$-$y$ plane)
and thus require less conformational distortion (i.e. reduction in entropy) in overlapping 
configurations. The variation of the free energy barrier height, $\Delta F$,
is shown in the inset of the figure on a log-log scale. Clearly, there is a power law relation
between $\Delta F$ and $D$, and a fit to the data yields $\Delta F \sim D^{-1.93\pm 0.01}$.
A slight deviation from this scaling is evident for the widest tube considered ($D=9$)
and is most likely due to chain-end effects.

To understand the observed dependence of $\Delta F$ on $N$ and $D$,
we consider an approach followed in Ref.~\onlinecite{racko2013segregation}, which
borrows from a previous treatment of a ring polymer in a confining channel.\cite{jung2012ring}
In Ref.~\onlinecite{jung2012ring}, it was suggested that a ring polymer confined in a cylinder can 
be viewed as two overlapping chains, each with a length that is half of that of the ring. 
The latter picture corresponds to the present system with $\lambda=0$. Polymers that
overlap can be viewed as occupying an effective tube with half the cross-sectional area,
i.e. with a reduced diameter of $D/\sqrt{2}$. In the context of the de~Gennes blob model, 
the confinement free energy of a single chain in a tube of diameter $D$
is given by $F_{\rm c}(N,D) \sim k_{\rm B}T N D^{-1/\nu}$, where the scaling exponent has
the value $\nu\approx 0.588$ for good solvent conditions. Two non-interacting chains
in the same tube have a free energy of confinement that is double this value. Thus,
the predicted free energy difference between two confined polymers of length $N$ in the
non-overlapping (high $\lambda$) and completely overlapping ($\lambda=0$) limits is thus,
$\Delta F = 2F_{\rm c}(N,D) - 2F_{\rm c}(N,D/\sqrt{2})$, which yields
$\Delta F  \sim k_{\rm B}T N D^{-\gamma}$, where $\gamma=1/\nu\approx 1.7$.
The predicted linear variation with respect to $N$ is consistent with the results
shown in Fig.~\ref{fig:F.scale}. On the other hand, the exponent of $\gamma=1.7$
differs from the value of $1.93\pm 0.01$ obtained from the data in Fig.~\ref{fig:F.N200}.
This discrepancy may arise in part from the simplistic approach of using an effective reduced
tube diameter to account for overlapping chains. More likely, it results from the 
use of the blob model outside the appropriate regime. Such discrepancies have been
observed in other studies. For example, Kim {\it et al.} have noted that the
scaling of the effective chain elasticity yields results consistent with the predictions
of the blob model only for relatively large tube diameters ($D>10$), while it yields
unexpected scaling for smaller diameters, such as those employed in this study.%
\cite{kim2013elasticity} More relevant to this work, Ra{\v{c}}ko and Cifra measured 
the confinement free energy of single $N=800$ chain of variable bending stiffness 
and observed a scaling exponent for the fully flexible case that clearly satisfies 
$\gamma > 1/\nu$ for tube diameters comparable those here (see Fig.~2 of 
Ref.~\onlinecite{racko2013segregation}).

\begin{figure}[!ht]
\begin{center}
\includegraphics[width=0.4\textwidth]{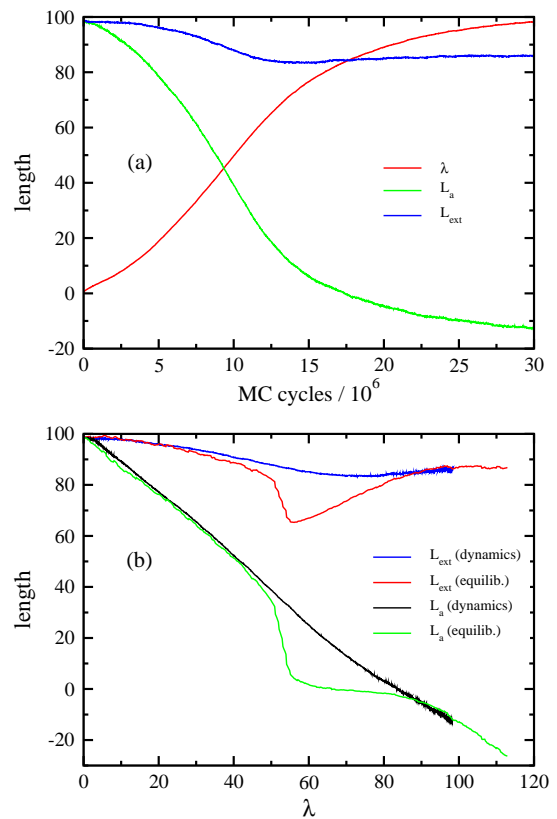}
\end{center}
\caption{(a) Variation of $\lambda$, $L_{\rm a}$ and $L_{\rm ext}$ with time during 
polymer segregation for a system with $N=200$ and $D=4$. The polymers start with 
$\lambda=0$.  (b) Variation of $L_{\rm a}$ and $L_{\rm ext}$ with $\lambda$. Results
are shown for segregation dynamics simulations and for equilibrium simulations.
}
\label{fig:Ldyn.time.N200.R2.5.Linfin}
\end{figure}

Let us now consider the dynamics of segregation for polymers confined to infinite length
cylinders. Figure~\ref{fig:Ldyn.time.N200.R2.5.Linfin}(a) shows the variation of the
average $\lambda$, $L_{\rm a}$ and $L_{\rm ext}$ with time, upon release from a state where the polymer
centers of mass overlap (i.e. $\lambda=0$). Results are shown for chains of length $N=200$
in a cylinder of diameter $D=4$. As expected, during segregation, $\lambda$ increases
and the overlap distance $L_{\rm a}$ decreases. Note that negative values of
$L_{\rm a}$ correspond to the (negative) distance between nearest monomers on the chains
after they no longer overlap. The polymer extension length $L_{\rm ext}$ initially decreases,
which is followed by a slight, gradual increase.  It is useful to consider the relation 
between the lengths $L_{\rm a}$ and $L_{\rm ext}$, and the center of mass distance, $\lambda$,
during segregation.
Figure~\ref{fig:Ldyn.time.N200.R2.5.Linfin}(b) shows this relation measured for the dynamics
simulations and compares the values with those from the (equilibrium) free energy 
calculations. In the initial stages of segregation, the dynamics and equilibrium 
curves are in reasonably good quantitative agreement.  At a later stage there is a more
pronounced discrepancy between the two sets of results for both $L_{\rm a}$ and $L_{\rm ext}$.
This occurs in region~III where the polymers are compressed and
in contact, but not overlapping. During the segregation process in the dynamics simulation,
the non-overlapping section of the polymer is not able to retract to its equilibrium
length, and thus $L_{\rm a}$ and $L_{\rm ext}$ are both greater 
than the equilibrium values. This out-of-equilibrium behavior eventually ends
at later times for separation distances, $\lambda$, corresponding to non-touching
chains. 

The discrepancy between the results for equilibrium and dynamics simulations
shown in Fig.~\ref{fig:Ldyn.time.N200.R2.5.Linfin}(b) was also observed for polymers 
of various lengths and cylinder diameters (data not shown).  Thus, out-of-equilibrium 
conformational behavior at later times appears to be a general result. Thus, 
using free energy functions to provide quantitative predictions of polymer segregation dynamics
during the later stages of the process is not appropriate.  This applies, for example, to using 
$F(\lambda)$ in an analysis of the Fokker-Planck equation to predict segregation times for large 
$L$ systems. 

\begin{figure}[!ht]
\begin{center}
\includegraphics[width=0.4\textwidth]{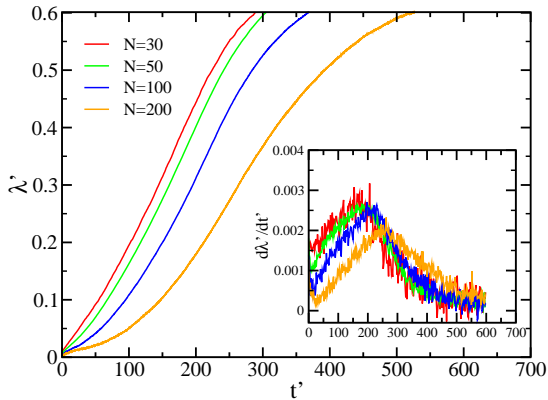}
\end{center}
\caption{
Variation of $\lambda'\equiv \lambda/N$ vs $t'\equiv t/N^2$ for two polymers in an infinitely 
long cylinder of diameter $D=2$.  Results are shown for different $N$. The inset shows the 
derivative $d\lambda'/dt'$ vs $t'$ for the same data. }
\label{fig:Lcm_dyn.time.Nscale.R1.5.Linfin}
\end{figure}

In the early stages of segregation the polymers do remain conformationally
relaxed. This corresponds to the regime where the polymers overlap (region~II, as 
labeled in Fig.~\ref{fig:F.Lab.N200.R2.5}), and the free energy varies approximately 
linearly with $\lambda$ for sufficiently long polymers. Thus, the magnitude of the 
effective force, $f\equiv |dF/d\lambda|$, is constant. The average relative velocity 
is approximately $v (\equiv d\lambda/dt) =2f/\gamma$, where $\gamma$ is the friction 
of each polymer and the factor of two accounts for the presence of two chains moving 
in opposite directions.  We assume that the friction of the individual monomers is 
additive, $\gamma\approx\gamma_0N$, Defining scaled variables $\lambda'\equiv \lambda/N$ 
and $t'\equiv t/N^2$, it follows that $d\lambda'/dt'$ should be constant and thus 
independent of $N$.  From Fig.~\ref{fig:F.scale} the regime of validity is 
$0.03\lesssim\lambda'\lesssim 0.4$ for $D=2$.  Figure~\ref{fig:Lcm_dyn.time.Nscale.R1.5.Linfin} 
shows $\lambda'$ vs $t'$ for systems of various $N$ for this tube diameter. The curves are 
approximately parallel to each other, with the higher $N$ curves shifted to higher $t'$.
To account for this shift, we use an approach similar to one employed in 
Ref.~\onlinecite{arnold2007time}. We first note that $F$ is approximately independent of 
$\lambda$ in regime~I ($\lambda'\lesssim 0.03$), within which polymer separation is 
governed by diffusion. The time required to reach the regime transition scales as 
$t_{\rm diff}\sim (0.03 L_{\rm a}^{(\rm max)})^2/2D_{\rm diff}\sim N^3$, where 
$L_{\rm a}^{(\rm max)}$~($\sim N$) is the extension length of the two fully overlapping 
polymers, and $D_{\rm diff}~(\sim 1/N)$ is the polymer diffusion coefficient. Thus, polymers 
are expected to cross into regime~II at a scaled time $t'\sim N$, which is approximately 
consistent with the observed shift in the curves.  In regime~II, the functions are not quite
linear but are rather slightly curved, which is due to the slight curvature in the free energy curves in 
Fig.~\ref{fig:Ldyn.time.N200.R2.5.Linfin}(a) in this regime.  As is evident from the inset 
of Fig.~\ref{fig:Lcm_dyn.time.Nscale.R1.5.Linfin}, the derivatives $d\lambda'/dt'$ 
are quantitatively comparable, though there 
is a slight reduction in the values with larger $N$. This is likely a consequence of the 
fact that scaled free energy functions in Fig.~\ref{fig:F.scale}(a)
have not quite converged over the range $N=30$--$200$. It may also be related to the very slight 
deviation from conformational equilibrium in regime~II evident for $L_{\rm a}$ in 
Fig.~\ref{fig:Ldyn.time.N200.R2.5.Linfin}. In spite of these discrepancies, the segregation 
dynamical behavior during the initial stages of the process is nevertheless reasonably
consistent with predictions using the free energy functions. 

\subsection{Finite length confinement cylinder}
\label{subsec:capped}

\begin{figure}[!ht]
\begin{center}
\includegraphics[width=0.4\textwidth]{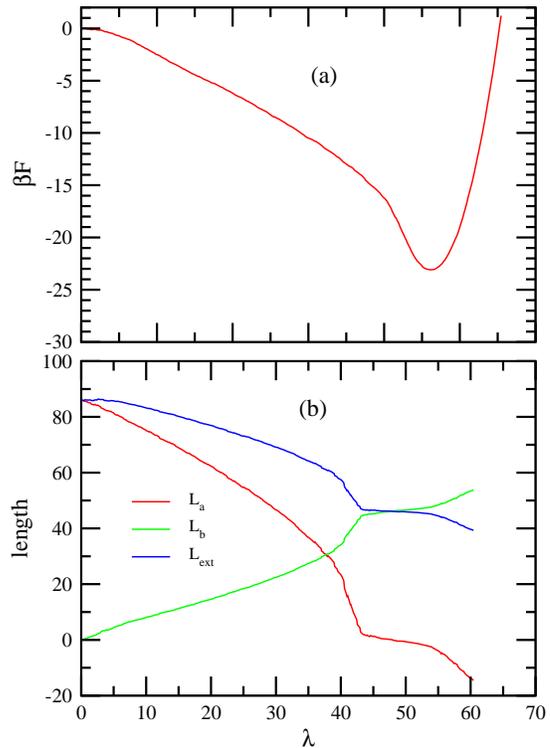}
\end{center}
\caption{(a) Free energy function for two chains of length $N=200$ in a
cylinder of diameter $D=4$ and length $L=90$.
(b) Variation of $L_{\rm a}$, $L_{\rm b}$ and $L_{\rm ext}$ with
$\lambda$.  Negative values of $L_{\rm a}$ correspond to the (negative)
distance between the nearest ends of non-overlapping chains.
}
\label{fig:F.Lab.N200.R2.5.L90}
\end{figure}

Next we consider the case of flexible chains ($\kappa=0$) confined to a cylindrical
tube of finite length, $L$. Figure~\ref{fig:F.Lab.N200.R2.5.L90}(a) shows the variation
of $F$ with $\lambda$ for a system of chains with length $N=200$ in a tube of diameter
$D=4$ and a length of $L=90$. Figure~\ref{fig:F.Lab.N200.R2.5.L90}(b) shows the variation
in the mean values of $L_{\rm a}$, $L_{\rm b}$, and $L_{\rm ext}$ with $\lambda$. 
The general trends are similar to those of the infinite-length case shown in 
Fig.~\ref{fig:F.Lab.N200.R2.5}, except for a steep rise in $F$ at high $\lambda$, which leads
to a free energy minimum at $\lambda \approx 57$. This rapid rise is due to the presence 
of the hemispheric caps on the confining tube. When the polymer center of mass separation 
is sufficiently great, the chains press against the hemisphere walls, and the crowding
of monomers reduces the conformational entropy of the chain. This crowding also leads
to a reduction in the polymer extension $L_{\rm ext}$ at large $\lambda$.

\begin{figure}[!ht]
\begin{center}
\includegraphics[width=0.4\textwidth]{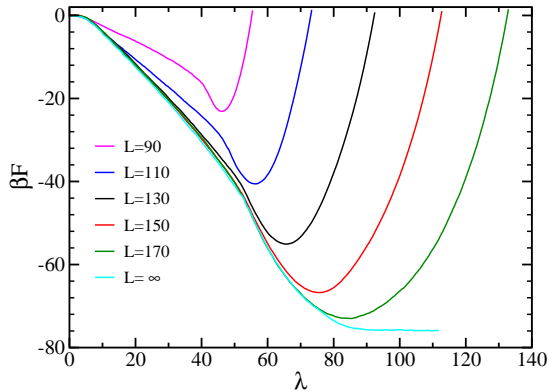}
\end{center}
\caption{Free energy functions for polymer chains of length $N=200$, confined
in a cylindrical tube of diameter $D=4$.  Results are shown for several different values
of the cylinder length, $L$. The free energy curves are shifted such that $F=0$ at
$\lambda=0$.}
\label{fig:F.L.N200.R2.5}
\end{figure}

Figure~\ref{fig:F.L.N200.R2.5} shows free energy functions for $N$=200 chains in a tube
of diameter $D$=4 for lengths ranging from $L$=90 to $L$=$\infty$. The curves have been
shifted along the vertical axis so that $F=0$ for complete overlap at $\lambda=0$.
The general features are similar for all of the finite-$L$ curves: each curve 
rises steeply at high $\lambda$ and exhibits a free energy minimum corresponding to the
most probable polymer center of mass separation distance in equilibrium.  
The height of the free energy barrier, $\Delta F\equiv F(0) - F(\lambda_{\rm min})$, 
and the location of the free energy minimum, $\lambda_{\rm min}$, both increase with
increasing $L$. In addition, the ratio $\lambda_{\rm min}/L\approx 0.51$ is nearly invariant 
with respect to tube length $L$ (data not shown). For larger $L$, where the polymers have 
negligible contact at distances where they are in contact with the end caps of the tube, 
the curves in Fig.~\ref{fig:F.L.N200.R2.5} exhibit a broad flat minimum in $F$
with a value equal to that for $L$=$\infty$ (data not shown). 

Some quantitative aspects of the results in Fig.~\ref{fig:F.L.N200.R2.5} can be understood 
by comparing the results with those of  $L=\infty$ shown in Fig.~\ref{fig:F.Lab.N200.R2.5}.
The free energy barrier height, $\Delta F$, for finite $L$ approaches that for
$L$=$\infty$  at a tube length equal to the minimum cylinder length for which two
polymers can fit inside without overlap or compression. This can be estimated
from the relation $L+D=2L_{\rm ext}(\infty)$, where $L_{\rm ext}(\infty)$ is average extension
for a single polymer in an infinite cylinder.  We note that $L_{\rm ext}(\infty)\approx 88$
for $N=200$ and $D=4$ (see the curve for $L_{\rm ext}$ at large $\lambda$
in Fig.~\ref{fig:F.Lab.N200.R2.5}(b)). Consequently, the critical tube length is
predicted to be $L\approx 172$, which is approximately consistent with the results
in the figure. Likewise, the convergence of $F$ for finite $L$ with that of $L=\infty$ in regime~II
up to $\lambda\approx 51$ for $L\gtrsim 130$ is also straightforward. At $\lambda=51$, the combined
average extension length of the overlapping chains for $L=\infty$ is $L_{\rm a}+2L_{\rm b}\approx 130$
(using data from Fig.~\ref{fig:F.Lab.N200.R2.5}(b)). This corresponds to the minimum length
of the confinement tube that can fit two polymers with this degree of overlap with no
deformation from longitudinal confinement.  More generally, the deviation of $F$ from the $L=\infty$ 
curve is expected to occur at lower degrees of overlap, i.e. lower $\lambda$, as the confinement 
length is reduced. This trend is clearly evident in Fig.~\ref{fig:F.L.N200.R2.5}.

\begin{figure}[!ht]
\begin{center}
\includegraphics[width=0.4\textwidth]{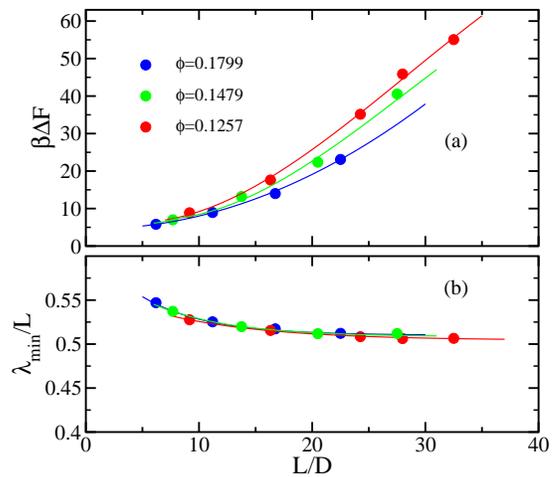}
\end{center}
\caption{(a) Free energy barrier height, $\beta \Delta F$, and (b) relative 
free energy minimum position, $\lambda_{\rm min}/L$, vs cylinder asymmetry, $L/D$. 
Data are shown for $N=200$ for three densities.  The solid lines are guides for the eye.
}
\label{fig:Fmin.N200}
\end{figure}

Next we consider the effects on the free energy of independently varying the confinement space aspect ratio, $L/D$,
and the monomer packing fraction, $\phi \equiv 2Nv/V$, where $v\equiv \pi\sigma^3/6$ is the volume per monomer 
and where $V$ is the (fixed) confinement volume.  Generally, increasing $L/D$ increases the free energy barrier, 
$\Delta F$, as well as the equilibrium separation, $\lambda_{\rm min}$.  Figure~\ref{fig:Fmin.N200}(a) shows 
the variation of $\Delta F$ with $L/D$ for three different packing fractions, and illustrates clearly the increase 
of $\Delta F$ with $L/D$.  It is also evident that the barrier height increases with decreasing $\phi$. Such 
an increase in the free energy cost of overlapping chains leads to more effective demixing. 
This effect may arise in part from the change in the width of the polymer relative to the tube dimensions
as $\phi$ is varied.  Nevertheless, our results are qualitatively 
consistent with the results of Jung {\it et al.} for a comparable model system, who found that polymer 
miscibility decreased (i.e. the tendency for chain demixing increased) with increasing $L/D$ and decreasing 
$\phi$.\cite{jung2012intrachain}
As $L/D$ becomes smaller, the barrier heights for different $\phi$ converge and tend toward zero. The
trend is consistent with the prediction of negligible free energy cost for overlapping chains under
spherical confinement.\cite{jun2007confined} The results are also qualitatively consistent with results 
from a recent simulation study for the concentrated regime.\cite{jacobsen2010demixing}
Figure~\ref{fig:Fmin.N200}(b) shows the relative free energy minimum position $\lambda_{\rm min}/L$ vs
$L/D$. Over the range of $L/D$ considered, $\lambda_{\rm min}/L$ is independent of $\phi$. In addition,
the dependence of $\lambda_{\rm min}/L$ on $L/D$ is very weak, and has a value near
$\lambda_{\rm min}/L\approx 0.51$. At low $L/D$, $\lambda_{\rm min}/L$ increases slightly, an indication
that the chains are becoming slightly more miscible. The trends for $\lambda_{\rm min}$  are consistent
with those for the positions of the peaks of the probability distribution shown in Fig.~2(a) of
Ref.~\onlinecite{jung2012intrachain}.

\begin{figure}[!ht]
\begin{center}
\includegraphics[width=0.4\textwidth]{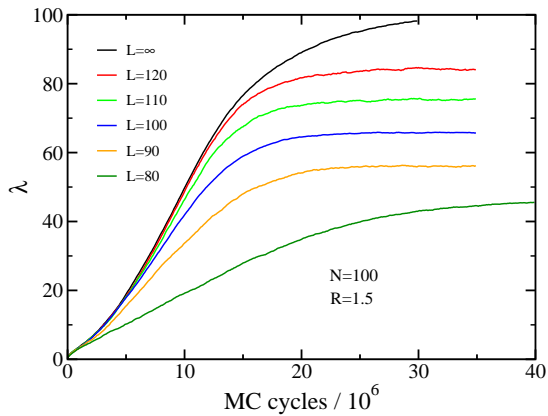}
\end{center}
\caption{Time dependence of $\lambda$ during segregation for polymers of length $N=200$ in 
a cylinder of diameter $D=4$ and finite length.  Results for several different cylinder lengths 
are shown.  
}
\label{fig:Lcm.cap.time.N200.R2.5}
\end{figure}

Figure~\ref{fig:Lcm.cap.time.N200.R2.5} shows the time-dependence of 
$\lambda$ during segregation for polymers of length $N=200$ in a tube
of diameter $D=4$. Data are shown for the same values of $L$ corresponding
to the free energy functions in Fig.~\ref{fig:F.L.N200.R2.5}. Generally, the segregation
rate increases monotonically with increasing $L$, though the rates are very close
for $L\leq 130$ at short times. Similar trends are observed for the time-dependence
of $L_{\rm a}$ during segregation (data not shown). These trends are consistent with 
expectations based on the trends in Fig.~\ref{fig:F.L.N200.R2.5} for the entropic force 
$f\equiv |dF/d\lambda|$, where we note that $f$ increases with increasing $L$, though 
$f$ converges for low $\lambda$ for $L\lesssim 130$.

\begin{figure}[!ht]
\begin{center}
\includegraphics[width=0.4\textwidth]{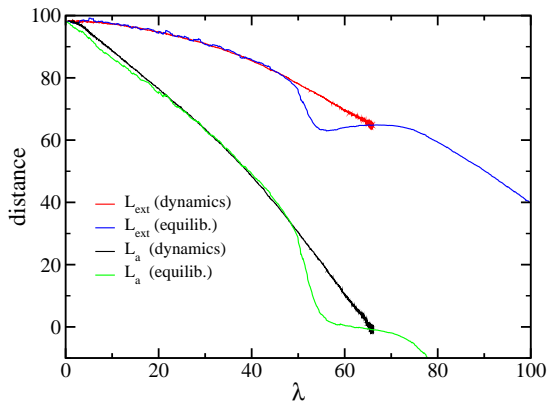}
\end{center}
\caption{Variation of $L_{\rm a}$ and $L_{\rm ext}$ with $\lambda$ during polymer segregation 
for a system with $N=200$, $D=4$, and $L=130$.  Results are shown for segregation dynamics 
simulations and for equilibrium simulations.
}
\label{fig:Ldyn.N200.R2.5.L130}
\end{figure}

Figure~\ref{fig:Ldyn.N200.R2.5.L130} shows the relationship of $L_{\rm a}$ and $L_{\rm ext}$
with $\lambda$ during segregation. The results are overlaid on those for a system
in equilibrium acquired during the simulations used to obtain the free energy functions.
The results follow a trend similar to that for $L=\infty$ evident in Fig.~\ref{fig:Ldyn.time.N200.R2.5.Linfin}.
During the first stage of segregation ($\lambda \lesssim 50$ and $t\leq 1.2\times 10^7$)
the polymer extension length $L_{\rm ext}$ is very close to the equilibrium values.
The overlap length, $L_{\rm a}$, is also close, though somewhat lower at very early times.
As in the $L=\infty$ case, the rapid reduction in $L_{\rm ext}$ and $L_{\rm a}$ for the
equilibrium system is not observed in the dynamics simulations. Thus, the system goes
out of equilibrium during this stage, before reaching the free energy minimum at $\lambda\approx 66$.

\subsection{Effects of bending rigidity}
\label{subsec:rigidity}

\begin{figure}[!ht]
\begin{center}
\includegraphics[width=0.4\textwidth]{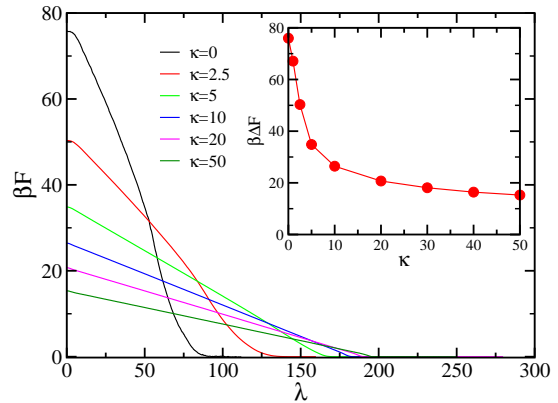}
\end{center}
\caption{Free energy functions for various degrees of bending rigidity, $\kappa$, 
for $N=200$ and $D=4$. The inset shows the variation of the free energy barrier
height with $\kappa$.
}
\label{fig:F.kappa.N200.R2.5}
\end{figure}

Finally, we consider the case of polymer chains of finite bending rigidity.
Figure~\ref{fig:F.kappa.N200.R2.5} shows several free energy functions for various 
values of the polymer stiffness constant, $\kappa$, for a system of chains of length
$N=200$ in an infinitely long confinement cylinder of diameter $D=4$. As the chains
become more rigid, the free energy barrier height, $\Delta F\equiv F(0)-F(\infty)$,
decreases. As is evident from the inset of the figure, $\Delta F$ appears to 
level off asymptotically to a finite value at large $\kappa$, in the limit
where the polymers become stiff rods. In addition to this trend, the free
energy functions become broader with increasing $\kappa$. This is due to 
an increase in the average extension length of the chains as they stiffen.
As chains extend further along the cylinder, they can interact
over greater distances. Thus, the reduction in conformational entropy (and
increase in $F$) due to this interaction will occur at larger $\lambda$.  For these $N=200$
chains, it is expected that the maximum value for which $F>0$ should
asymptotically approach $\lambda=200$, which is the maximum separation of
rigid linear chains for which contact between the chains is possible.
(This estimate neglects the fluctuations in the contour length due
to the slight variability of the bond length.)
This is consistent with the results in the figure.

The trends observed in Fig.~\ref{fig:F.kappa.N200.R2.5} are helpful for understanding 
some results of the recent study by Ra{\v{c}}ko and Cifra.\cite{racko2013segregation} 
In that study, Langevin dynamics simulations were used to study the effects of
bending rigidity on the segregation rate for polymers under cylindrical confinement.
It was found that flexible chains segregate significantly faster than less 
flexible chains.  They observed a fast-segregation regime for low persistence length $P$, for 
which the segregation rate decreased rapidly with increasing stiffness, and a slow-segregation 
regime at higher $P$, for which there was little variation of the rate with chain stiffness. 
For sufficiently high bending rigidity, the chains even failed to segregate over the 
time scale of the simulation.  The transition between the regimes occurred when 
$P\sim D$. Fast segregation is expected to correspond to a large effective driving force 
magnitude, $f\equiv |dF/d\lambda|$, which our results show clearly decreases with 
$\kappa$ (and, thus, $P$). In addition, the driving force is also closely correlated 
with the barrier height, $\Delta F$, which, as the inset of the figure shows, decreases 
abruptly until $P\approx 5$. Note that $\kappa/k_{\rm B}T=P/\langle l_{\rm bond}\rangle$, 
where $\langle l_{\rm bond}\rangle\approx \sigma=1$ is
the mean bond length. Thus, the transition occurs when $P\sim D$, and the trends in 
segregation rates predicted from the free energy results are consistent with those of the 
dynamics simulations.

\begin{figure}[!ht]
\begin{center}
\includegraphics[width=0.4\textwidth]{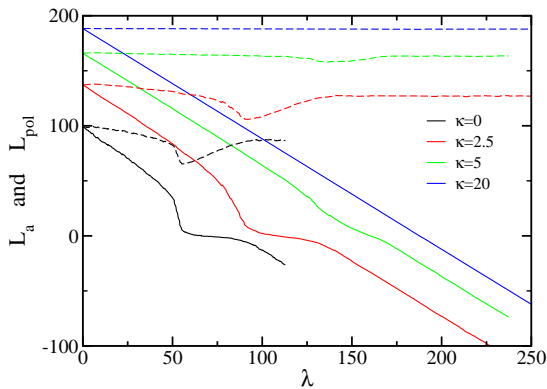}
\end{center}
\caption{Overlap length $L_{\rm a}$ (solid lines) and polymer extension length $L_{\rm ext}$
(dashed lines) vs $\lambda$ for various degrees of bending rigidity, $\kappa$,
for $N=200$ and $D=4$. The data were generated using the same simulations used
for the results in Fig.~\ref{fig:F.kappa.N200.R2.5}.
}
\label{fig:Lab.kappa.N200.R2.5}
\end{figure}

Figure~\ref{fig:Lab.kappa.N200.R2.5} shows the overlap distance $L_{\rm a}$ and
extension length $L_{\rm ext}$ as a function of $\lambda$ for various $\kappa$
These data were generated using the same free energy simulations used for
Fig.~\ref{fig:F.kappa.N200.R2.5}. As expected, stiffening the chains leads
to greater longitudinal extension of the polymers. This, in turn, leads to
a greater degree of overlap for any given separation, $\lambda$. Another
interesting trend is the gradual disappearance of regime~III, in which
$L_{\rm a}\approx 0$ and the extension length $L_{\rm ext}$ first rapidly 
decreases due to polymer retraction and then slowly increases with $\lambda$.
As $\kappa$ is increased, the variation of $L_{\rm a}$ with $\lambda$ 
becomes more uniformly linear. In addition, the ``retraction dip'' in $L_{\rm ext}$ 
disappears, and the difference in $L_{\rm ext}$ between the overlap and non-overlapping 
regions vanishes.  As in Fig.~\ref{fig:F.kappa.N200.R2.5}, the cross-over between 
the regimes occurs when $P\sim D$.

\begin{figure}[!ht]
\begin{center}
\includegraphics[width=0.4\textwidth]{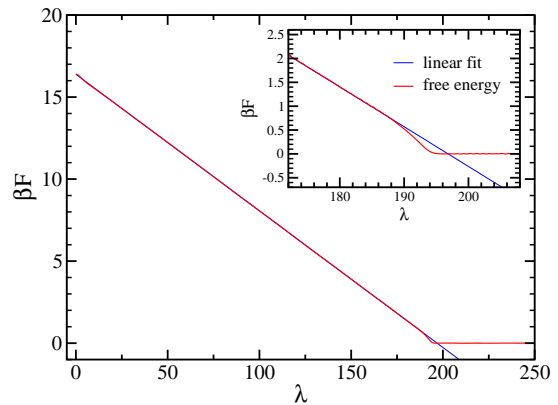}
\end{center}
\caption{Free energy function for polymer chains of length $N=200$ and bending
rigidity $\kappa=40$ in an infinite tube of diameter $D=4$. The blue curve shows
a linear fit to most of the data. The inset shows a close up near the point 
past which the polymers no longer overlap.
}
\label{fig:F.kappa.N200.R3}
\end{figure}

Consider the case for chains of length $N=200$ and bending stiffness $\kappa=40$ 
in a $D=4$ cylinder. Since $P\approx 40$ and the contour length of the
chain, $C$, has a value $C\approx 200$, the condition for the Odijk regime, 
$D \ll P \ll C$ is marginally satisfied.
Figure~\ref{fig:F.kappa.N200.R3} shows the free energy function for this system.
The curve is highly linear over the range $12\lesssim\lambda\lesssim 188$, with
slight deviations from linearity for $0\leq \lambda \lesssim 12$ and 
$188\lesssim \lambda \lesssim 195$. (The inset of the figure illustrates the
deviation for the latter case.) These regions correspond to $L_{b} \lesssim 12$
and $L_{\rm a}\lesssim 7$, respectively. 

\begin{figure}[!ht]
\begin{center}
\includegraphics[width=0.4\textwidth]{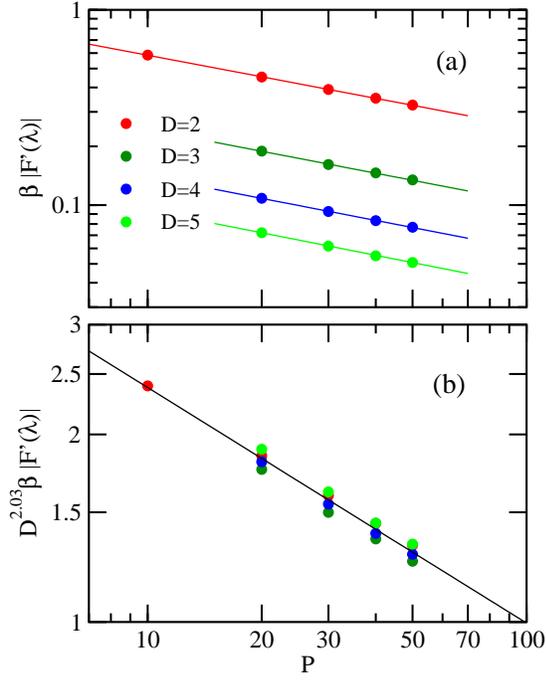}
\end{center}
\caption{(a) Magnitude of the free energy function derivative, $\beta |dF/d\lambda|$, vs
polymer persistence length, $P$. The derivative is obtained from a fit to the 
linear region of $F(\lambda)$ for systems that marginally satisfy the condition for the 
Odijk regime.  Data are shown for $N=200$ chains for four cylinder diameters and are
chosen to satisfy $4D\leq P \leq 4C$.  The solid lines are power law fits of the four data sets.  
(b) Scaled free energy function derivatives, $D^{2.03}\beta |F^\prime(\lambda)|$ vs 
$P$ using the data in (a). The solid line is a power law fit to all of the data.  The fit 
yielded $D^{2.03}\beta |F^\prime(\lambda)|\sim P^{-0.37}$. }
\label{fig:Fderiv}
\end{figure}

The trend where $F$ varies linearly with $\lambda$, with only slight discrepancies
near the edges of the regime where the chains overlap, is a generic feature of
systems we have examined that marginally satisfy the Odijk condition. The derivative 
$F^\prime(\lambda)$ in the linear domain was calculated for these systems by fitting 
$F(\lambda)$ in this region with a linear function.
Figure~\ref{fig:Fderiv}(a) shows the variation of the magnitude of the derivative,
$\beta |F'(\lambda)|$, vs $P$ for four different cylinder diameters.  
For each $D$, a fit to the function $\beta |F'| \sim P^{-\alpha}$, yields
exponents near $\alpha\approx 0.37$. Figure~\ref{fig:Fderiv}(b) shows 
the same data, except with the derivatives scaled by a factor of $D^{2.03}$. This
scaling factor was found to yield the best collapse of the data onto a single
function. Thus, the results suggest that $\beta |F'| \sim D^{-\beta} P^{-\alpha}$
in the Odijk regime, where $\beta\approx 2$ and $\alpha\approx 0.37$.  

A theoretical prediction for the scaling of $F'(\lambda)$ with respect to $D$ and $P$ 
in the Odijk regime can be derived as follows.  First recall that the confinement 
free energy of a single chain, $F_1$, can be approximated by assigning $k_{\rm B}T$ for each 
deflection of the chain off of the walls of the confining cylinder. Thus, 
$F_1\sim k_{\rm B}T L_{\rm ext}/l_{\rm def}$, where the Odijk deflection length is given 
by $l_{\rm def}\approx D^{2/3}P^{1/3}$.\cite{odijk1983statistics} The confinement free energy of 
a system of two overlapping chains can be written, $F_{\rm c} = 2F_1 + F^{(\rm int)}$, where
$F^{\rm (int)}$ is the interaction free energy associated with the overlapping
portions of the chains. This term can be estimated using an approximation of the
free energy of a system of long, hard rigid rods. For that system, recall that 
the interaction free energy per particle, in the 2nd virial approximation, is given by 
$F^{\rm (int)}/Nk_{\rm B}T = (l^2\sigma N/V) \langle|\sin\gamma|\rangle$ for 
$N$ rods of length $l$ and diameter $\sigma$ confined to a volume $V$,
where $\gamma$ is the angle between two rods.\cite{odijk1986theory} For highly
aligned systems, $\langle|\sin\gamma|\rangle \sim \sqrt{\langle\theta^2\rangle}$,
where $\theta$ is the angle between the rod and the alignment direction.
To apply this result to the present system, we treat a deflection segment of
length $l_{\rm d}$ as a rigid rod, and note that $V$ corresponds to the volume
over which intermolecular segments overlap. Thus, we substitute
$l\rightarrow l_{\rm d}\sim D^{2/3}P^{1/3}$, $N\rightarrow 2N_{\rm d}\equiv 2L/l_{\rm d}$,
and $V\sim L_{\rm a} D^2$. In addition, assuming that the alignment arises principally from
the confinement, $\langle\theta^2\rangle\sim (D/l_{\rm d})^2$. It follows that
the overlap free energy is
$\beta F \equiv \beta(F_{\rm c}-2F_1) \sim  L_{\rm a}\sigma D^{-5/3} P^{-1/3}$.  Next, we note that 
$L_{\rm a}\approx L_{\rm ext}-\lambda$, where the extension length $L_{\rm ext}$  was 
confirmed to be constant with respect to $\lambda$ (data not shown). This leads finally to
the prediction that $d(\beta F)/d\lambda \sim -\sigma D^{-\beta}P^{-\alpha}$,
where $\beta=5/3$ and $\alpha=1/3$. 

The predicted exponents are comparable to the measured exponents, though the differences
between them are still significant. It is likely that the ratios $P/D$ and $C/P$ are
not sufficiently large for some of the approximations to be completely valid.
The deviations of this scaling from the form $\beta F^\prime(\lambda) \sim D^{-\beta}P^{-\alpha}$, 
evident in the scatter of the data points about the best fit curve in Fig.~\ref{fig:Fderiv}(b),
also suggest that the scaling regime has not been reached. We speculate that agreement with 
the predicted scaling would improve if the Odijk conditions were better satisfied. 
However, such verification requires using much longer polymer chains, which 
is not feasible for us at present. 

\section{Conclusions}
\label{sec:conclusions}

In this study, we have used MC simulations to systematically study the overlap free energy 
function, $F(\lambda)$, for two linear polymers with center of mass separation, $\lambda$, confined 
in a cylindrical tube of diameter $D$ and length $L$. As expected, $F$ increases with decreasing
polymer separation. In the case of flexible chains confined to
tubes of infinite length, the functions are  characterized by well defined domains. 
One unexpected result is the existence of a domain in which the polymers do not overlap, 
but are in contact and are compressed. This domain disappears when the polymers stiffen,
with the cross-over at $P\sim D$, where $P$ is the persistence length. We observe
scaling laws for the free energy functions for both flexible and semi-flexible chain systems
that are qualitatively consistent with theoretical predictions using scaling
arguments. However, there are deviations are observed between the measured and predicted 
values of the scaling exponents, which are most likely due to insufficiently satisfying
the conditions defining the de~Gennes and Odijk regimes.  For confinement cylinders of 
finite $L$, the free energy barrier height increases with increasing
confinement aspect ratio, $L/D$, at fixed packing fraction, $\phi$, and decreases with
increasing $\phi$ at fixed $L/D$. These results are consistent with previous measurements
of polymer miscibility for a comparable model system.\cite{jung2012intrachain}

While the main focus of this work is the characterization of the free energy functions, 
we also carried out simulations to study the dynamics of segregation. The goal here was
to examine the relationship between the segregation rates and the free energy functions. 
One notable result is the deviation from conformational quasi-equilibrium observed at later 
times during polymer separation. This implies that the free energy functions cannot be used 
(e.g. in the Fokker-Plank equation) to make quantitatively accurate predictions for the 
segregation dynamics, since the $F(\lambda)$ characterizes a system in equilibrium.  
Nevertheless, we did observe the intuitively appealing qualitative trend that the rate of 
polymer segregation is generally faster when $|dF/d\lambda|$ increases.  Furthermore, 
although we did not not study the segregation dynamics for chains of finite stiffness in 
this work, the observed effect of varying the chain rigidity on the free energy functions 
does qualitatively explain the observed trends in a recent Langevin dynamics simulation
study.\cite{racko2013segregation}

This work has focused on the overlap free energy for systems of relatively short ($N\leq 200$)
linear polymers. In future work, we will extend the present study in several ways.
In order to make the results more applicable to chromosome separation in bacteria,
we will examine the free energy functions of cylindrically confined ring polymers,
as well as examine the role of crowding agents, as in Ref.~\onlinecite{shin2014mixing}.
Preliminary calculations using ring polymers have thus far yielded trends comparable to 
those observed here. In addition, we will study systems of much longer chain lengths by
employing more efficient Monte Carlo simulation methods. This will allow a more effective
test of the scaling predictions described in this work. It will also enable us to examine
the changes in the free energy as a system of semi-flexible polymers transitions from the 
Odijk regime to the de~Gennes regime upon increasing the confinement diameter.

\begin{acknowledgments}
This work was supported by the Natural Sciences and Engineering Research Council of Canada (NSERC).
We are grateful to the Atlantic Computational Excellence Network (ACEnet) for use of
their computational resources.
\end{acknowledgments}

%

\begin{thebibliography}{36}%
\makeatletter
\providecommand \@ifxundefined [1]{%
 \@ifx{#1\undefined}
}%
\providecommand \@ifnum [1]{%
 \ifnum #1\expandafter \@firstoftwo
 \else \expandafter \@secondoftwo
 \fi
}%
\providecommand \@ifx [1]{%
 \ifx #1\expandafter \@firstoftwo
 \else \expandafter \@secondoftwo
 \fi
}%
\providecommand \natexlab [1]{#1}%
\providecommand \enquote  [1]{``#1''}%
\providecommand \bibnamefont  [1]{#1}%
\providecommand \bibfnamefont [1]{#1}%
\providecommand \citenamefont [1]{#1}%
\providecommand \href@noop [0]{\@secondoftwo}%
\providecommand \href [0]{\begingroup \@sanitize@url \@href}%
\providecommand \@href[1]{\@@startlink{#1}\@@href}%
\providecommand \@@href[1]{\endgroup#1\@@endlink}%
\providecommand \@sanitize@url [0]{\catcode `\\12\catcode `\$12\catcode
  `\&12\catcode `\#12\catcode `\^12\catcode `\_12\catcode `\%12\relax}%
\providecommand \@@startlink[1]{}%
\providecommand \@@endlink[0]{}%
\providecommand \url  [0]{\begingroup\@sanitize@url \@url }%
\providecommand \@url [1]{\endgroup\@href {#1}{\urlprefix }}%
\providecommand \urlprefix  [0]{URL }%
\providecommand \Eprint [0]{\href }%
\providecommand \doibase [0]{http://dx.doi.org/}%
\providecommand \selectlanguage [0]{\@gobble}%
\providecommand \bibinfo  [0]{\@secondoftwo}%
\providecommand \bibfield  [0]{\@secondoftwo}%
\providecommand \translation [1]{[#1]}%
\providecommand \BibitemOpen [0]{}%
\providecommand \bibitemStop [0]{}%
\providecommand \bibitemNoStop [0]{.\EOS\space}%
\providecommand \EOS [0]{\spacefactor3000\relax}%
\providecommand \BibitemShut  [1]{\csname bibitem#1\endcsname}%
\let\auto@bib@innerbib\@empty
\bibitem [{\citenamefont {Grosberg}\ \emph {et~al.}(1982)\citenamefont
  {Grosberg}, \citenamefont {Khalatur},\ and\ \citenamefont
  {Khokhlov}}]{grosberg1982polymeric}%
  \BibitemOpen
  \bibfield  {author} {\bibinfo {author} {\bibfnamefont {A.~Y.}\ \bibnamefont
  {Grosberg}}, \bibinfo {author} {\bibfnamefont {P.~G.}\ \bibnamefont
  {Khalatur}}, \ and\ \bibinfo {author} {\bibfnamefont {A.~R.}\ \bibnamefont
  {Khokhlov}},\ }\href@noop {} {\bibfield  {journal} {\bibinfo  {journal}
  {Makromol. Chem. Rapid Comm.}\ }\textbf {\bibinfo {volume} {3}},\ \bibinfo
  {pages} {709} (\bibinfo {year} {1982})}\BibitemShut {NoStop}%
\bibitem [{\citenamefont {Dautenhahn}\ and\ \citenamefont
  {Hall}(1994)}]{dautenhahn1994monte}%
  \BibitemOpen
  \bibfield  {author} {\bibinfo {author} {\bibfnamefont {J.}~\bibnamefont
  {Dautenhahn}}\ and\ \bibinfo {author} {\bibfnamefont {C.~K.}\ \bibnamefont
  {Hall}},\ }\href@noop {} {\bibfield  {journal} {\bibinfo  {journal}
  {Macromolecules}\ }\textbf {\bibinfo {volume} {27}},\ \bibinfo {pages} {5399}
  (\bibinfo {year} {1994})}\BibitemShut {NoStop}%
\bibitem [{\citenamefont {Daoud}\ and\ \citenamefont
  {De~Gennes}(1977)}]{daoud1977statistics}%
  \BibitemOpen
  \bibfield  {author} {\bibinfo {author} {\bibfnamefont {M.}~\bibnamefont
  {Daoud}}\ and\ \bibinfo {author} {\bibfnamefont {P.}~\bibnamefont
  {De~Gennes}},\ }\href@noop {} {\bibfield  {journal} {\bibinfo  {journal} {J.
  Phys. (Paris)}\ }\textbf {\bibinfo {volume} {38}},\ \bibinfo {pages} {85}
  (\bibinfo {year} {1977})}\BibitemShut {NoStop}%
\bibitem [{\citenamefont {Jun}\ and\ \citenamefont
  {Mulder}(2006)}]{jun2006entropy}%
  \BibitemOpen
  \bibfield  {author} {\bibinfo {author} {\bibfnamefont {S.}~\bibnamefont
  {Jun}}\ and\ \bibinfo {author} {\bibfnamefont {B.}~\bibnamefont {Mulder}},\
  }\href@noop {} {\bibfield  {journal} {\bibinfo  {journal} {P. Natl. Acad.
  Sci. USA}\ }\textbf {\bibinfo {volume} {103}},\ \bibinfo {pages} {12388}
  (\bibinfo {year} {2006})}\BibitemShut {NoStop}%
\bibitem [{\citenamefont {Jun}\ and\ \citenamefont
  {Wright}(2010)}]{jun2010entropy}%
  \BibitemOpen
  \bibfield  {author} {\bibinfo {author} {\bibfnamefont {S.}~\bibnamefont
  {Jun}}\ and\ \bibinfo {author} {\bibfnamefont {A.}~\bibnamefont {Wright}},\
  }\href@noop {} {\bibfield  {journal} {\bibinfo  {journal} {Nat. Rev.
  Microbiol.}\ }\textbf {\bibinfo {volume} {8}},\ \bibinfo {pages} {600}
  (\bibinfo {year} {2010})}\BibitemShut {NoStop}%
\bibitem [{\citenamefont {Youngren}\ \emph {et~al.}(2014)\citenamefont
  {Youngren}, \citenamefont {Nielsen}, \citenamefont {Jun},\ and\ \citenamefont
  {Austin}}]{youngren2014multifork}%
  \BibitemOpen
  \bibfield  {author} {\bibinfo {author} {\bibfnamefont {B.}~\bibnamefont
  {Youngren}}, \bibinfo {author} {\bibfnamefont {H.~J.}\ \bibnamefont
  {Nielsen}}, \bibinfo {author} {\bibfnamefont {S.}~\bibnamefont {Jun}}, \ and\
  \bibinfo {author} {\bibfnamefont {S.}~\bibnamefont {Austin}},\ }\href@noop {}
  {\bibfield  {journal} {\bibinfo  {journal} {Genes \& development}\ }\textbf
  {\bibinfo {volume} {28}},\ \bibinfo {pages} {71} (\bibinfo {year}
  {2014})}\BibitemShut {NoStop}%
\bibitem [{\citenamefont {Le~Chat}\ and\ \citenamefont
  {Esp{\'e}li}(2012)}]{lechat2012let}%
  \BibitemOpen
  \bibfield  {author} {\bibinfo {author} {\bibfnamefont {L.}~\bibnamefont
  {Le~Chat}}\ and\ \bibinfo {author} {\bibfnamefont {O.}~\bibnamefont
  {Esp{\'e}li}},\ }\href@noop {} {\bibfield  {journal} {\bibinfo  {journal}
  {Mol. Microbiol.}\ }\textbf {\bibinfo {volume} {86}},\ \bibinfo {pages}
  {1285} (\bibinfo {year} {2012})}\BibitemShut {NoStop}%
\bibitem [{\citenamefont {Yazdi}\ \emph {et~al.}(2012)\citenamefont {Yazdi},
  \citenamefont {Guet}, \citenamefont {Johnson},\ and\ \citenamefont
  {Marko}}]{yazdi2012variation}%
  \BibitemOpen
  \bibfield  {author} {\bibinfo {author} {\bibfnamefont {N.~H.}\ \bibnamefont
  {Yazdi}}, \bibinfo {author} {\bibfnamefont {C.~C.}\ \bibnamefont {Guet}},
  \bibinfo {author} {\bibfnamefont {R.~C.}\ \bibnamefont {Johnson}}, \ and\
  \bibinfo {author} {\bibfnamefont {J.~F.}\ \bibnamefont {Marko}},\ }\href@noop
  {} {\bibfield  {journal} {\bibinfo  {journal} {Mol. Microbiol.}\ }\textbf
  {\bibinfo {volume} {86}},\ \bibinfo {pages} {1318} (\bibinfo {year}
  {2012})}\BibitemShut {NoStop}%
\bibitem [{\citenamefont {Kuwada}\ \emph {et~al.}(2013)\citenamefont {Kuwada},
  \citenamefont {Cheveralls}, \citenamefont {Traxler},\ and\ \citenamefont
  {Wiggins}}]{kuwada2013mapping}%
  \BibitemOpen
  \bibfield  {author} {\bibinfo {author} {\bibfnamefont {N.~J.}\ \bibnamefont
  {Kuwada}}, \bibinfo {author} {\bibfnamefont {K.~C.}\ \bibnamefont
  {Cheveralls}}, \bibinfo {author} {\bibfnamefont {B.}~\bibnamefont {Traxler}},
  \ and\ \bibinfo {author} {\bibfnamefont {P.~A.}\ \bibnamefont {Wiggins}},\
  }\href@noop {} {\bibfield  {journal} {\bibinfo  {journal} {Nucleic Acids
  Res.}\ }\textbf {\bibinfo {volume} {41}},\ \bibinfo {pages} {7370} (\bibinfo
  {year} {2013})}\BibitemShut {NoStop}%
\bibitem [{\citenamefont {Di~Ventura}\ \emph {et~al.}(2013)\citenamefont
  {Di~Ventura}, \citenamefont {Knecht}, \citenamefont {Andreas}, \citenamefont
  {Godinez}, \citenamefont {Fritsche}, \citenamefont {Rohr}, \citenamefont
  {Nickel}, \citenamefont {Heermann},\ and\ \citenamefont
  {Sourjik}}]{diventura2013chromosome}%
  \BibitemOpen
  \bibfield  {author} {\bibinfo {author} {\bibfnamefont {B.}~\bibnamefont
  {Di~Ventura}}, \bibinfo {author} {\bibfnamefont {B.}~\bibnamefont {Knecht}},
  \bibinfo {author} {\bibfnamefont {H.}~\bibnamefont {Andreas}}, \bibinfo
  {author} {\bibfnamefont {W.~J.}\ \bibnamefont {Godinez}}, \bibinfo {author}
  {\bibfnamefont {M.}~\bibnamefont {Fritsche}}, \bibinfo {author}
  {\bibfnamefont {K.}~\bibnamefont {Rohr}}, \bibinfo {author} {\bibfnamefont
  {W.}~\bibnamefont {Nickel}}, \bibinfo {author} {\bibfnamefont {D.~W.}\
  \bibnamefont {Heermann}}, \ and\ \bibinfo {author} {\bibfnamefont
  {V.}~\bibnamefont {Sourjik}},\ }\href@noop {} {\bibfield  {journal} {\bibinfo
   {journal} {Mol. Syst. Biol.}\ }\textbf {\bibinfo {volume} {9}} (\bibinfo
  {year} {2013})}\BibitemShut {NoStop}%
\bibitem [{\citenamefont {Junier}\ \emph {et~al.}(2014)\citenamefont {Junier},
  \citenamefont {Boccard},\ and\ \citenamefont
  {Esp{\'e}li}}]{junier2014polymer}%
  \BibitemOpen
  \bibfield  {author} {\bibinfo {author} {\bibfnamefont {I.}~\bibnamefont
  {Junier}}, \bibinfo {author} {\bibfnamefont {F.}~\bibnamefont {Boccard}}, \
  and\ \bibinfo {author} {\bibfnamefont {O.}~\bibnamefont {Esp{\'e}li}},\
  }\href@noop {} {\bibfield  {journal} {\bibinfo  {journal} {Nucleic Acids
  Res.}\ }\textbf {\bibinfo {volume} {42}},\ \bibinfo {pages} {1461} (\bibinfo
  {year} {2014})}\BibitemShut {NoStop}%
\bibitem [{\citenamefont {Reyes-Lamothe}\ \emph {et~al.}(2012)\citenamefont
  {Reyes-Lamothe}, \citenamefont {Nicolas},\ and\ \citenamefont
  {Sherratt}}]{reyes2012chromosome}%
  \BibitemOpen
  \bibfield  {author} {\bibinfo {author} {\bibfnamefont {R.}~\bibnamefont
  {Reyes-Lamothe}}, \bibinfo {author} {\bibfnamefont {E.}~\bibnamefont
  {Nicolas}}, \ and\ \bibinfo {author} {\bibfnamefont {D.~J.}\ \bibnamefont
  {Sherratt}},\ }\href@noop {} {\bibfield  {journal} {\bibinfo  {journal}
  {Annu. Rev. Genet.}\ }\textbf {\bibinfo {volume} {46}},\ \bibinfo {pages}
  {121} (\bibinfo {year} {2012})}\BibitemShut {NoStop}%
\bibitem [{\citenamefont {Wang}\ \emph {et~al.}(2013)\citenamefont {Wang},
  \citenamefont {Llopis},\ and\ \citenamefont {Rudner}}]{wang2013organization}%
  \BibitemOpen
  \bibfield  {author} {\bibinfo {author} {\bibfnamefont {X.}~\bibnamefont
  {Wang}}, \bibinfo {author} {\bibfnamefont {P.~M.}\ \bibnamefont {Llopis}}, \
  and\ \bibinfo {author} {\bibfnamefont {D.~Z.}\ \bibnamefont {Rudner}},\
  }\href@noop {} {\bibfield  {journal} {\bibinfo  {journal} {Nat. Rev. Genet.}\
  }\textbf {\bibinfo {volume} {14}},\ \bibinfo {pages} {191} (\bibinfo {year}
  {2013})}\BibitemShut {NoStop}%
\bibitem [{\citenamefont {Teraoka}\ and\ \citenamefont
  {Wang}(2004)}]{teraoka2004computer}%
  \BibitemOpen
  \bibfield  {author} {\bibinfo {author} {\bibfnamefont {I.}~\bibnamefont
  {Teraoka}}\ and\ \bibinfo {author} {\bibfnamefont {Y.}~\bibnamefont {Wang}},\
  }\href@noop {} {\bibfield  {journal} {\bibinfo  {journal} {Polymer}\ }\textbf
  {\bibinfo {volume} {45}},\ \bibinfo {pages} {3835} (\bibinfo {year}
  {2004})}\BibitemShut {NoStop}%
\bibitem [{\citenamefont {Jun}\ \emph {et~al.}(2007)\citenamefont {Jun},
  \citenamefont {Arnold},\ and\ \citenamefont {Ha}}]{jun2007confined}%
  \BibitemOpen
  \bibfield  {author} {\bibinfo {author} {\bibfnamefont {S.}~\bibnamefont
  {Jun}}, \bibinfo {author} {\bibfnamefont {A.}~\bibnamefont {Arnold}}, \ and\
  \bibinfo {author} {\bibfnamefont {B.-Y.}\ \bibnamefont {Ha}},\ }\href@noop {}
  {\bibfield  {journal} {\bibinfo  {journal} {Phys. Rev. Lett.}\ }\textbf
  {\bibinfo {volume} {98}},\ \bibinfo {pages} {128303} (\bibinfo {year}
  {2007})}\BibitemShut {NoStop}%
\bibitem [{\citenamefont {Arnold}\ and\ \citenamefont
  {Jun}(2007)}]{arnold2007time}%
  \BibitemOpen
  \bibfield  {author} {\bibinfo {author} {\bibfnamefont {A.}~\bibnamefont
  {Arnold}}\ and\ \bibinfo {author} {\bibfnamefont {S.}~\bibnamefont {Jun}},\
  }\href@noop {} {\bibfield  {journal} {\bibinfo  {journal} {Phys. Rev. E}\
  }\textbf {\bibinfo {volume} {76}},\ \bibinfo {pages} {031901} (\bibinfo
  {year} {2007})}\BibitemShut {NoStop}%
\bibitem [{\citenamefont {Jacobsen}(2010)}]{jacobsen2010demixing}%
  \BibitemOpen
  \bibfield  {author} {\bibinfo {author} {\bibfnamefont {J.~L.}\ \bibnamefont
  {Jacobsen}},\ }\href@noop {} {\bibfield  {journal} {\bibinfo  {journal}
  {Phys. Rev. E}\ }\textbf {\bibinfo {volume} {82}},\ \bibinfo {pages} {051802}
  (\bibinfo {year} {2010})}\BibitemShut {NoStop}%
\bibitem [{\citenamefont {Jung}\ and\ \citenamefont
  {Ha}(2010)}]{jung2010overlapping}%
  \BibitemOpen
  \bibfield  {author} {\bibinfo {author} {\bibfnamefont {Y.}~\bibnamefont
  {Jung}}\ and\ \bibinfo {author} {\bibfnamefont {B.-Y.}\ \bibnamefont {Ha}},\
  }\href@noop {} {\bibfield  {journal} {\bibinfo  {journal} {Phys. Rev. E}\
  }\textbf {\bibinfo {volume} {82}},\ \bibinfo {pages} {051926} (\bibinfo
  {year} {2010})}\BibitemShut {NoStop}%
\bibitem [{\citenamefont {Jung}\ \emph
  {et~al.}(2012{\natexlab{a}})\citenamefont {Jung}, \citenamefont {Jeon},
  \citenamefont {Kim}, \citenamefont {Jeong}, \citenamefont {Jun},\ and\
  \citenamefont {Ha}}]{jung2012ring}%
  \BibitemOpen
  \bibfield  {author} {\bibinfo {author} {\bibfnamefont {Y.}~\bibnamefont
  {Jung}}, \bibinfo {author} {\bibfnamefont {C.}~\bibnamefont {Jeon}}, \bibinfo
  {author} {\bibfnamefont {J.}~\bibnamefont {Kim}}, \bibinfo {author}
  {\bibfnamefont {H.}~\bibnamefont {Jeong}}, \bibinfo {author} {\bibfnamefont
  {S.}~\bibnamefont {Jun}}, \ and\ \bibinfo {author} {\bibfnamefont {B.-Y.}\
  \bibnamefont {Ha}},\ }\href@noop {} {\bibfield  {journal} {\bibinfo
  {journal} {Soft Matter}\ }\textbf {\bibinfo {volume} {8}},\ \bibinfo {pages}
  {2095} (\bibinfo {year} {2012}{\natexlab{a}})}\BibitemShut {NoStop}%
\bibitem [{\citenamefont {Jung}\ \emph
  {et~al.}(2012{\natexlab{b}})\citenamefont {Jung}, \citenamefont {Kim},
  \citenamefont {Jun},\ and\ \citenamefont {Ha}}]{jung2012intrachain}%
  \BibitemOpen
  \bibfield  {author} {\bibinfo {author} {\bibfnamefont {Y.}~\bibnamefont
  {Jung}}, \bibinfo {author} {\bibfnamefont {J.}~\bibnamefont {Kim}}, \bibinfo
  {author} {\bibfnamefont {S.}~\bibnamefont {Jun}}, \ and\ \bibinfo {author}
  {\bibfnamefont {B.-Y.}\ \bibnamefont {Ha}},\ }\href@noop {} {\bibfield
  {journal} {\bibinfo  {journal} {Macromolecules}\ }\textbf {\bibinfo {volume}
  {45}},\ \bibinfo {pages} {3256} (\bibinfo {year}
  {2012}{\natexlab{b}})}\BibitemShut {NoStop}%
\bibitem [{\citenamefont {Liu}\ and\ \citenamefont
  {Chakraborty}(2012)}]{liu2012segregation}%
  \BibitemOpen
  \bibfield  {author} {\bibinfo {author} {\bibfnamefont {Y.}~\bibnamefont
  {Liu}}\ and\ \bibinfo {author} {\bibfnamefont {B.}~\bibnamefont
  {Chakraborty}},\ }\href@noop {} {\bibfield  {journal} {\bibinfo  {journal}
  {Phys. Biol.}\ }\textbf {\bibinfo {volume} {9}},\ \bibinfo {pages} {066005}
  (\bibinfo {year} {2012})}\BibitemShut {NoStop}%
\bibitem [{\citenamefont {Dorier}\ and\ \citenamefont
  {Stasiak}(2013)}]{dorier2013modelling}%
  \BibitemOpen
  \bibfield  {author} {\bibinfo {author} {\bibfnamefont {J.}~\bibnamefont
  {Dorier}}\ and\ \bibinfo {author} {\bibfnamefont {A.}~\bibnamefont
  {Stasiak}},\ }\href@noop {} {\bibfield  {journal} {\bibinfo  {journal}
  {Nucleic Acids Res.}\ }\textbf {\bibinfo {volume} {41}},\ \bibinfo {pages}
  {6808} (\bibinfo {year} {2013})}\BibitemShut {NoStop}%
\bibitem [{\citenamefont {Ra{\v{c}}ko}\ and\ \citenamefont
  {Cifra}(2013)}]{racko2013segregation}%
  \BibitemOpen
  \bibfield  {author} {\bibinfo {author} {\bibfnamefont {D.}~\bibnamefont
  {Ra{\v{c}}ko}}\ and\ \bibinfo {author} {\bibfnamefont {P.}~\bibnamefont
  {Cifra}},\ }\href@noop {} {\bibfield  {journal} {\bibinfo  {journal} {J.
  Chem. Phys.}\ }\textbf {\bibinfo {volume} {138}},\ \bibinfo {pages} {184904}
  (\bibinfo {year} {2013})}\BibitemShut {NoStop}%
\bibitem [{\citenamefont {Shin}\ \emph {et~al.}(2014)\citenamefont {Shin},
  \citenamefont {Cherstvy},\ and\ \citenamefont {Metzler}}]{shin2014mixing}%
  \BibitemOpen
  \bibfield  {author} {\bibinfo {author} {\bibfnamefont {J.}~\bibnamefont
  {Shin}}, \bibinfo {author} {\bibfnamefont {A.~G.}\ \bibnamefont {Cherstvy}},
  \ and\ \bibinfo {author} {\bibfnamefont {R.}~\bibnamefont {Metzler}},\
  }\href@noop {} {\bibfield  {journal} {\bibinfo  {journal} {New J. Phys.}\
  }\textbf {\bibinfo {volume} {16}},\ \bibinfo {pages} {053047} (\bibinfo
  {year} {2014})}\BibitemShut {NoStop}%
\bibitem [{\citenamefont {Milchev}\ \emph {et~al.}(2014)\citenamefont
  {Milchev}, \citenamefont {M{\"{u}}ller},\ and\ \citenamefont
  {Klushin}}]{milchev2014arm}%
  \BibitemOpen
  \bibfield  {author} {\bibinfo {author} {\bibfnamefont {A.}~\bibnamefont
  {Milchev}}, \bibinfo {author} {\bibfnamefont {M.}~\bibnamefont
  {M{\"{u}}ller}}, \ and\ \bibinfo {author} {\bibfnamefont {L.}~\bibnamefont
  {Klushin}},\ }\href@noop {} {\bibfield  {journal} {\bibinfo  {journal}
  {Macromolecules}\ }\textbf {\bibinfo {volume} {47}},\ \bibinfo {pages} {2156}
  (\bibinfo {year} {2014})}\BibitemShut {NoStop}%
\bibitem [{\citenamefont {Frenkel}\ and\ \citenamefont
  {Smit}(2002)}]{frenkel2002understanding}%
  \BibitemOpen
  \bibfield  {author} {\bibinfo {author} {\bibfnamefont {D.}~\bibnamefont
  {Frenkel}}\ and\ \bibinfo {author} {\bibfnamefont {B.}~\bibnamefont {Smit}},\
  }\href@noop {} {\emph {\bibinfo {title} {Understanding Molecular Simulation:
  From Algorithms to Applications}}},\ \bibinfo {edition} {2nd}\ ed.\ (\bibinfo
   {publisher} {Academic Press},\ \bibinfo {address} {London},\ \bibinfo {year}
  {2002})\ Chap.~\bibinfo {chapter} {7}\BibitemShut {NoStop}%
\bibitem [{\citenamefont {Polson}\ \emph {et~al.}(2013)\citenamefont {Polson},
  \citenamefont {Hassanabad},\ and\ \citenamefont
  {McCaffrey}}]{polson2013simulation}%
  \BibitemOpen
  \bibfield  {author} {\bibinfo {author} {\bibfnamefont {J.~M.}\ \bibnamefont
  {Polson}}, \bibinfo {author} {\bibfnamefont {M.~F.}\ \bibnamefont
  {Hassanabad}}, \ and\ \bibinfo {author} {\bibfnamefont {A.}~\bibnamefont
  {McCaffrey}},\ }\href@noop {} {\bibfield  {journal} {\bibinfo  {journal} {J.
  Chem. Phys.}\ }\textbf {\bibinfo {volume} {138}},\ \bibinfo {pages} {024906}
  (\bibinfo {year} {2013})}\BibitemShut {NoStop}%
\bibitem [{\citenamefont {Polson}\ and\ \citenamefont
  {McCaffrey}(2013)}]{polson2013polymer}%
  \BibitemOpen
  \bibfield  {author} {\bibinfo {author} {\bibfnamefont {J.~M.}\ \bibnamefont
  {Polson}}\ and\ \bibinfo {author} {\bibfnamefont {A.~C.}\ \bibnamefont
  {McCaffrey}},\ }\href@noop {} {\bibfield  {journal} {\bibinfo  {journal} {J.
  Chem. Phys.}\ }\textbf {\bibinfo {volume} {138}},\ \bibinfo {pages} {174902}
  (\bibinfo {year} {2013})}\BibitemShut {NoStop}%
\bibitem [{\citenamefont {Polson}\ and\ \citenamefont
  {Dunn}(2014)}]{polson2014evaluating}%
  \BibitemOpen
  \bibfield  {author} {\bibinfo {author} {\bibfnamefont {J.~M.}\ \bibnamefont
  {Polson}}\ and\ \bibinfo {author} {\bibfnamefont {T.~R.}\ \bibnamefont
  {Dunn}},\ }\href@noop {} {\bibfield  {journal} {\bibinfo  {journal} {J. Chem.
  Phys.}\ }\textbf {\bibinfo {volume} {140}},\ \bibinfo {pages} {184904}
  (\bibinfo {year} {2014})}\BibitemShut {NoStop}%
\bibitem [{not()}]{note1}%
  \BibitemOpen
  \href@noop {} {}\bibinfo {note} {The study of
  Ref.~\onlinecite{milchev2014arm} also examined the variation of the free
  energy with polymer separation in a cylinder. In that case, however, the
  polymers were both tethered to a capped end of the tube. In the present
  study, by contrast, the polymers move freely in the cylinder and much higher
  free energy barriers are observed.}\BibitemShut {Stop}%
\bibitem [{\citenamefont {Narros}\ \emph {et~al.}(2010)\citenamefont {Narros},
  \citenamefont {Moreno},\ and\ \citenamefont {Likos}}]{narros2010influence}%
  \BibitemOpen
  \bibfield  {author} {\bibinfo {author} {\bibfnamefont {A.}~\bibnamefont
  {Narros}}, \bibinfo {author} {\bibfnamefont {A.~J.}\ \bibnamefont {Moreno}},
  \ and\ \bibinfo {author} {\bibfnamefont {C.~N.}\ \bibnamefont {Likos}},\
  }\href@noop {} {\bibfield  {journal} {\bibinfo  {journal} {Soft Matter}\
  }\textbf {\bibinfo {volume} {6}},\ \bibinfo {pages} {2435} (\bibinfo {year}
  {2010})}\BibitemShut {NoStop}%
\bibitem [{\citenamefont {Narros}\ \emph {et~al.}(2013)\citenamefont {Narros},
  \citenamefont {Moreno},\ and\ \citenamefont
  {Likos}}]{narros2013architecture}%
  \BibitemOpen
  \bibfield  {author} {\bibinfo {author} {\bibfnamefont {A.}~\bibnamefont
  {Narros}}, \bibinfo {author} {\bibfnamefont {A.~J.}\ \bibnamefont {Moreno}},
  \ and\ \bibinfo {author} {\bibfnamefont {C.~N.}\ \bibnamefont {Likos}},\
  }\href@noop {} {\bibfield  {journal} {\bibinfo  {journal} {Macromolecules}\
  }\textbf {\bibinfo {volume} {46}},\ \bibinfo {pages} {9437} (\bibinfo {year}
  {2013})}\BibitemShut {NoStop}%
\bibitem [{\citenamefont {Jun}\ \emph {et~al.}(2008)\citenamefont {Jun},
  \citenamefont {Thirumalai},\ and\ \citenamefont {Ha}}]{jun2008compression}%
  \BibitemOpen
  \bibfield  {author} {\bibinfo {author} {\bibfnamefont {S.}~\bibnamefont
  {Jun}}, \bibinfo {author} {\bibfnamefont {D.}~\bibnamefont {Thirumalai}}, \
  and\ \bibinfo {author} {\bibfnamefont {B.-Y.}\ \bibnamefont {Ha}},\
  }\href@noop {} {\bibfield  {journal} {\bibinfo  {journal} {Phys. Rev. Lett.}\
  }\textbf {\bibinfo {volume} {101}},\ \bibinfo {pages} {138101} (\bibinfo
  {year} {2008})}\BibitemShut {NoStop}%
\bibitem [{\citenamefont {Kim}\ \emph {et~al.}(2013)\citenamefont {Kim},
  \citenamefont {Jeon}, \citenamefont {Jeong}, \citenamefont {Jung},\ and\
  \citenamefont {Ha}}]{kim2013elasticity}%
  \BibitemOpen
  \bibfield  {author} {\bibinfo {author} {\bibfnamefont {J.}~\bibnamefont
  {Kim}}, \bibinfo {author} {\bibfnamefont {C.}~\bibnamefont {Jeon}}, \bibinfo
  {author} {\bibfnamefont {H.}~\bibnamefont {Jeong}}, \bibinfo {author}
  {\bibfnamefont {Y.}~\bibnamefont {Jung}}, \ and\ \bibinfo {author}
  {\bibfnamefont {B.-Y.}\ \bibnamefont {Ha}},\ }\href@noop {} {\bibfield
  {journal} {\bibinfo  {journal} {Soft Matter}\ }\textbf {\bibinfo {volume}
  {9}},\ \bibinfo {pages} {6142} (\bibinfo {year} {2013})}\BibitemShut
  {NoStop}%
\bibitem [{\citenamefont {Odijk}(1983)}]{odijk1983statistics}%
  \BibitemOpen
  \bibfield  {author} {\bibinfo {author} {\bibfnamefont {T.}~\bibnamefont
  {Odijk}},\ }\href@noop {} {\bibfield  {journal} {\bibinfo  {journal}
  {Macromolecules}\ }\textbf {\bibinfo {volume} {16}},\ \bibinfo {pages} {1340}
  (\bibinfo {year} {1983})}\BibitemShut {NoStop}%
\bibitem [{\citenamefont {Odijk}(1986)}]{odijk1986theory}%
  \BibitemOpen
  \bibfield  {author} {\bibinfo {author} {\bibfnamefont {T.}~\bibnamefont
  {Odijk}},\ }\href@noop {} {\bibfield  {journal} {\bibinfo  {journal}
  {Macromolecules}\ }\textbf {\bibinfo {volume} {19}},\ \bibinfo {pages} {2313}
  (\bibinfo {year} {1986})}\BibitemShut {NoStop}%
\end{thebibliography}
%

%

\end{document}